\RequirePackage{filecontents}

\documentclass[preprint, 12pt]{elsarticle}
\usepackage[T1]{fontenc}
\usepackage[utf8]{inputenc} 
\usepackage{geometry}
\usepackage{xurl}
\usepackage{soul}
\usepackage[final,nopatch=footnote]{microtype}
\usepackage{graphicx}
\usepackage{subcaption}
\usepackage{color}
\usepackage{microtype}
\usepackage{comment}
\usepackage{epsfig}
\usepackage{array}
\usepackage{multirow}
\usepackage{booktabs}
\usepackage{fancyhdr}
\usepackage{amsmath}
\usepackage{amssymb}
\usepackage{amstext}
\usepackage{hyperref}
\usepackage{adjustbox}
\usepackage{indentfirst}
\hypersetup{
     colorlinks=true,
     linkcolor=blue,
     filecolor=blue,
     citecolor = blue,      
     urlcolor=blue,
     }

\usepackage{framed} 
\usepackage{multicol} 
\usepackage{nomencl} 
\makenomenclature
\renewcommand*\nompreamble{\begin{multicols}{2}}
\renewcommand*\nompostamble{\end{multicols}}

\biboptions{sort&compress}
\DeclareUnicodeCharacter{2212}{-}
\usepackage[skip = 5pt, font = footnotesize, labelfont = bf, belowskip=-5pt,]{caption}
\usepackage{float}

\newcommand{\sulfate}{0.1 M Na\textsubscript{2}SO\textsubscript{4(aq.)} solution adjusted to pH 4}
\newcommand{\icrit}{\textit{i\textsubscript{crit}}}
\newcommand{\ipass}{\textit{i\textsubscript{pass}}}
\newcommand{\zimag}{$-$Z\textsubscript{imag}}
\newcommand{\zmod}{$|Z|$}

\journal{Acta Materialia}
\begin{document}
\begin{frontmatter}

\title{\textmd{Revisiting the Influence of Sn in Cu-Al alloys: A Third Element Effect Enabling Stainless Steel Type Aqueous Passivation Behavior}}

\author[1,2]{Debashish Sur \corref{cor1}}
\author[3]{Nathan C. Smith}
\author[4]{Elaf A. Anber}
\author[1,2]{Kaitlyn L. Anderson}
\author[1,2]{Peter F. Connors}
\author[4]{Daniel Foley}
\author[4]{Mitra L. Taheri}
\author[5]{Junsoo Han}
\author[3]{Christopher M. Wolverton}
\author[1,2]{John R. Scully}

\cortext[cor1]{Corresponding author: ds8vw@virginia.edu, sur.debashish96@gmail.com}

\affiliation[1]{organization={Center for Electrochemical Science and Engineering},
           addressline={University of Virginia}, 
            city={Charlottesville},
            postcode={22903}, 
            state={VA},
            country={USA}}
\affiliation[2]{organization={Department of Materials Science and Engineering},
            addressline={University of Virginia}, 
            city={Charlottesville},
            postcode={22903}, 
            state={VA},
            country={USA}}
\affiliation[3]{organization={Department of Materials Science and Engineering},
            addressline={Northwestern University}, 
            city={Evanston},
            postcode={602808}, 
            state={IL},
            country={USA}}
\affiliation[4]{organization={Department of Materials Science and Engineering}, 
addressline = {Johns Hopkins University}, 
city = {Baltimore}, 
state = {MD}, 
postcode = {21218},
country = {USA}}
\affiliation[5]{organization={Laboratoire Interfaces et Systèmes Electrochimiques (LISE)},
            addressline={Sorbonne Université}, 
            city={Paris},
            postcode={75005}, 
            country={France}}

\date{}

\begin{abstract}
    \hspace{0.5 cm}The influence of Sn alloying additions on the aqueous passivation behavior of Cu-Al alloys was revisited and found to function as a new \textit{third element effect} in acidified 0.1 M $Na_2SO_4$ solution. The role of each element during the process of aqueous passivation was investigated using electrochemical and surface-sensitive \textit{ex-situ} and \textit{in-operando} spectroscopic techniques. The connection between passivation and the atomic arrangements of atoms in the solid solution was supported by first principles' based cluster expansion calculations and Monte Carlo simulations probing the chemical short-range order in the Cu-Al-Sn system. High purity Sn, like high purity Cu, did not passivate in the test environment, whereas high purity Al formed a passive film with a stable passive current density of 0.01 $mA\cdot cm^{-2}$. Cu-$x$Al-Sn solid solution alloys where $x>18$ at.\%, containing less than 3 at.\% Sn additions exhibited lower corrosion rates than Cu-$x$Al alloys, brought by Al(III) and Sn(IV, II) unidentified complex oxides formation on the surface. A strong influence of Sn on Al(III) passivation was observed, i.e., strongly suggesting a third element effect type behavior. Possible governing processes explaining the stainless steel type corrosion behavior are discussed, providing insights for exploring novel synergies in the design of corrosion resistant alloys. 
\end{abstract}


\end{frontmatter}

\begin{table*}[!tbt]   
\begin{framed}
\nomenclature[s001]{$a$}{Lattice constant}
\nomenclature[s002]{$\alpha^p_{ij}$}{Warren-Cowley CSRO parameter of atom species $i$ and $j$ within neighbor shell $p$}
\nomenclature[s003]{$i$}{Current density}
\nomenclature[s003]{\ipass }{Passive current density}
\nomenclature[s003]{\icrit }{Critical current density}
\nomenclature[s003]{$i_M$}{Spectral dissolution current density of metal $M$}
\nomenclature[s003]{$i_{sum}$}{Sum of all elemental dissolution current densities}
\nomenclature[s003]{$i^{cg}_M$ }{Congruent dissolution current density of metal $M$}
\nomenclature[s004]{$E$ }{Potential vs. SHE}
\nomenclature[s004]{$E_{corr}$ }{Corrosion potential}
\nomenclature[s004]{$h$}{Number of dissolved monolayers}
\nomenclature[s004]{$R_p$}{Polarization resistance from EIS}

\nomenclature[s005]{$Z_{imag}$}{Imaginary component of complex impedance Z}
\nomenclature[s005]{${\lvert Z\rvert}$}{Modulus of complex impedance Z}

\nomenclature[s011]{$\Theta^\prime _M$}
{Normalized surface enrichment}
\nomenclature[s004]{$f_M$}{XPS based surface Enrichment factor of metal $M$ }
\nomenclature[s004]{$X^{film}_M$}{XPS based surface mole fraction of cationic species of element M}
\nomenclature[s004]{$X^{bulk}_M$}{Bulk mole fraction of element M}

\nomenclature[s012]{AESEC}{Atomic Emission Spectroelectrochemistry}
\nomenclature[s012]{LSV}{Linear Sweep Voltammetery}
\nomenclature[s012]{EIS}{Electrochemical Impedance Spectroscopy}
\nomenclature[s012]{OCP}{Open Circuit Potential}
\nomenclature[s012]{SHE}{Standard Hydrogen Electrode}
\nomenclature[s012]{CA}{Chronoamperometry}
\nomenclature[s012]{XPS}{X-ray Photoelectron Spectroscropy}
\nomenclature[s012]{SEM}{Scanning Electron Microscopy}
\nomenclature[s012]{STEM}{Scanning Transmission Electron Microscopy}
\nomenclature[s012]{EDS}{Energy Dispersive Spectroscopy}
\nomenclature[s012]{XRD}{X-ray Diffraction}
\nomenclature[s012]{FIB}{Focussed Ion Beam}
\nomenclature[s012]{CSRO}{Chemical Short Range Ordering}
\nomenclature[s012]{CPE}{Constant Phase Element}
\nomenclature[s012]{CE}{Cluster Expansion}
\nomenclature[s012]{DFT}{Density Functional Theory}
\nomenclature[s012]{MC}{Monte Carlo simulations}
\nomenclature[s012]{TEE}{Third Element Effect}
\nomenclature[s012]{$nn$}{Nearest neighbor}
\printnomenclature
\end{framed}
\end{table*}

\pagebreak
\section{Introduction}

Copper (Cu) and its alloys are widely valued for their excellent corrosion resistance, combined with high electrical and thermal conductivity, ease of fabrication, and resistance to biofouling in marine environments. Cu exhibits minimal corrosion in clean air, water, and deaerated non-oxidizing acids. This excellent resistance is evidenced by Cu artifacts discovered in near-pristine condition after being buried underground for thousands of years. \cite{Cohen2005CorrosionAlloys, CopperDevelopmentAssociationInc.2024ApplicationAlloys} In structural applications, Cu and its alloys are generally regarded as corrosion-resistant, relying on protective cuprous oxide in mild atmospheres.\cite{Francis2010TheEngineers} In alkaline environments, the passive film consists of an outer layer of cupric oxy/hydroxides and cuprous oxide as the inner layer.\cite{Francis2010TheEngineers} However, in acidic, neutral, or sulfate-containing environments, the ability to maintain a stable passive film (thick) is significantly diminished, resulting in active corrosion and the development of thick, non-protective surface layers, commonly known as tarnish.\cite{Strandberg1998SomeChloride} Consequently, in the past few decades, there has been some ongoing search for stainless Cu-based alloys.\cite{Francis2010TheEngineers} Alloying with a passivating element like Nickel (Ni) has shown promise in diminishing Cu corrosion by promoting the formation of its oxy/hydroxide rich passive film.\cite{Francis2010TheEngineers, Hummel1988TheTechniques} 
Similar to Fe-Cr alloys \cite{Uhlig1952}, Cu alloys with thin passive films also demonstrate a threshold-type behavior with respect to Ni where a minimum amount of the passivating component in solid solution is required to achieve alloy passivity in acidified solutions.\cite{Mansfeld1969EffectAlloys} For example, more than 50 wt.\% Ni in solid solution with Cu is required to achieve $Ni(OH)_2$ rich passivity exposed to acidified sulfate and chloride solutions,\cite{Mansfeld1969EffectAlloys, Badawy2009TheSolutions, Beccaria1989DealloyingWater}, resulting in commercial Cu-Ni alloys such as C70600 (90 wt.\% Cu) and C71500 (70 wt.\% Ni). In contrast to Ni, Zn additions in Cu do not significantly enhance protection via passive film formation.\cite{Jinturkar1998DissolutionAlloys} However, Zn reduces Cu dissolution rates by dezincification, where Zn preferentially dissolves.\cite{Jinturkar1998DissolutionAlloys} While Zn is less effective than Ni towards corrosion, it improves mechanical properties. Consequently, several Cu-Ni-Zn based commercial alloys (e.g., C73500, Cu-18Ni-8Zn in wt.\%), known as \textit{Ni silvers}, are widely used in food and musical instruments industries due to their good corrosion resistance and aesthetically pleasing white metal appearance.

The alloying of Cu with Al has been found to improve its corrosion resistance while simultaneously enhancing mechanical strength achieved through solution strengthening and formation of brittle $\beta$ phase  ($Cu_3Al$ L1$_{2}$ type ordered structure) when the Al content exceeds 20 at.\%.\cite{SarangodeSouza2022EffectsSolution, Kuo2006TheSolution, Ahlers2003AgeingAlloys} Virtanen et al. showed that 11 at.\% Al addition in Cu forms a FCC solid solution that has shown an active-passive behavior during anodic linear sweep voltammetry (LSV) in aerated 0.1 N $Na_2SO_4$, under conditions where pure Cu extensively tarnishes. Further increasing the Al alloying content up to the solubility limit in copper of 19 at.\% lowered the critical (\icrit) and passive (\ipass) current densities by approximately 10 times, suggesting some alloy passivation with lowered active dissolution of the alloy, in this case mostly Cu.\cite{Virtanen1993PassivityAlloys} The passivation in these alloys can be attributed to the ability of Al to form $Al_2O_3$, which has a higher affinity for oxygen than Cu.\cite{Scully1990TheCorrosion} 
Commercially available alloys such as Ni-Al bronzes (Cu-9Al-4Ni (wt.\%)) and \textit{Nordic Gold} (Cu-5Al-5Zn-1Sn (wt.\%)) are used in various structural applications, including marine and heat exchangers, as well as minting coins due to their good corrosion resistance and mechanical strength compared to pure Cu. However, Cu-Al-Ni alloy containing $\approx 24$ at.\% Al still showed a relatively high {\icrit} value of 10 $mA\cdot cm^{-2}$ and relatively high {\ipass} value of 1 $mA\cdot cm^{-2}$ in sulfate solutions \cite{Kuo2006TheSolution, Amegroud2021APotential}, as well as in deaerated 3.5 wt.\% NaCl \cite{Gojic2011ElectrochemicalAlloy}, suggesting that just Al additions alone do not guarantee excellent passivation behavior.

Significant amounts of Sn (more than 10 wt.\%) in Cu form a solid solution alloy reported to be beneficial in lowering its corrosion rate in sulfate solutions \cite{Muller2013-CuSnProcesses}. According to most phase diagrams, the maximum solubility of Sn in Cu as a solid solution is approximately 15.8 wt.\%. In  0.01 M $Na_2SO_4$ solution, the {\icrit} value decreases for a Cu-13 wt.\% Sn alloy to approximately $0.3$ $mA\cdot cm^{-2}$, lower by an order of magnitude compared to pure Cu.\cite{Mabille2003MechanismConditions} Raman spectroscopy and scanning electrochemical microscopy (SEM) imaging of the different layers developed on the surface suggests Sn forms an early layer of Sn(IV) oxides in a mixture with Cu(I) oxide through which Cu atoms diffuse to form an outer Cu(II) oxy/hydroxide layer.\cite{Robbiola2008CharacterisationSolution} Hutchinson et al. reported minimal dissolution of Sn species from Cu-4.5Sn-0.1Zn and Cu-9.7Sn-0.1Zn (wt.\%) alloys during galvanostatic experiments in artificial perspiration solution using the atomic emission spectroelectrochemistry (AESEC) technique.\cite{Hutchison2018PatinaPerspiration}  This early formation of an inner layer of Sn species is suggested to create a blocking effect to Cu dissolution as observed on ancient bronzes \cite{Robbiola2008CharacterisationSolution, Robbiola1998MorphologyAlloys}, the protective ability of which depends on the Sn content in the alloy. Minimal differences have been reported for Cu-7Sn and Cu-11Sn (wt.\%), during anodic LSV in 0.01 M $Na_2SO_4$ solution,\cite{Muller2013-CuSnProcesses} where any beneficial effect from small Sn additions may show up in a more dilute sulfate solution and after a long exposure at corrosion potential as observed for a Cu-6(wt.\%)Sn alloy after 12 days of exposure at corrosion potential in 15 mM $Na_2SO_4$.\cite{Satovic2009CorrosionSolution}. Commercial tin bronzes (C90700) and leaded tin bronzes (C94500, C92200) are used in bearings, piston rings, and fittings for their good pressure tightness and corrosion resistance behavior.\cite{CopperDevelopmentAssociationInc.2024ApplicationAlloys}

Ternary Cu-Al-Sn alloys have been scarcely explored for their corrosion behavior in aqueous solutions. Virtanen et al. reported that Sn can enhance the stainless behavior of these alloys in aerated 0.1 M $Na_2SO_4$ solution, adjusted to different pH values.\cite{Virtanen1993PassivityAlloys} In this solution, Cu-10Al-5Sn (wt.\%) exhibited the smallest {\icrit}, {\ipass} values compared to Cu-10Al (wt.\%) and high purity Cu using anodic LSV experiments. Wojtas et al. reported that the benefit of Sn is a function of Al content when testing passivation behaviors of Cu-Al-Sn alloys containing a variation of Al content (0-20 at.\%) and Sn content (0-5 at.\% Sn) in the same solution.\cite{Wojtas1995ElectrochemicalAlloys} The {\icrit} of Cu-15Al (at.\%) decreased from $\approx 1$ $mA\cdot cm^{-2}$ to $\approx 0.05$ $mA\cdot cm^{-2}$ as 1 at.\% Sn was added. Further additions of Sn up to 3 at.\% only reduced the {\icrit} value to $\approx 0.05$ $mA\cdot cm^{-2}$. Increasing the Al content to 19 at.\% while maintaining 1 at.\% Sn showed a slight decrease in {\icrit} values and significantly lower {\ipass} values, which was still 100 times lower compared to the {\icrit} and {\ipass} values of Cu-19 Al (at.\%). The Al content was enriched in the surface layer up to 30 at.\% for the  Cu-19Al-3Sn (at.\%) after 1 hour and 60 at.\% after 60 hours of immersion at corrosion potential via Auger electron spectroscopy peak analysis. The authors argued that this enrichment of $Al_2O_3$ is the key reason behind low corrosion rates and that the contributions from Sn oxides are improbable due to very low Sn content in the bulk alloy.\cite{Wojtas1995ElectrochemicalAlloys}. These works suggest a strong Al-Sn beneficial synergy; however, even though the underlying mechanisms remain unresolved.

When Cr is added as a third element to a binary alloy such as Fe-Al, the threshold Al content required to achieve $Al_2O_3$ type passivity at elevated temperatures above 1000 $^\circ$C is reduced from approximately 16-18 at.\% to 3-6 at.\%. This synergy has been coined as the `\textit{third-element effect}' of Cr, where it acts as a facilitator for Al to form $Al_2O_3$.\cite{Tomaszewicz1978, Wagner1957FormationMetals, Wagner1965PassivityTemperatures} C. Wagner explored this in other systems, such as the Cu-Al-Zn, where a dense $Al_2O_3$ layer is formed at a considerably lower Al alloy content in the presence of Zn than in its absence.\cite{Wagner1965PassivityTemperatures} Without Zn, a minimum of $\approx 14$ at.\% Al is required to achieve $Al_2O_3$ rich passivation, whereas without Al, $\approx 23$ at.\% of Zn is required to observe $ZnO$ rich scale, which still shows a para-linear kinetics during oxidation. However, with 4 at.\% Al addition to Cu-22Zn (at.\%), a significantly lower mass gain with parabolic kinetics was observed.\cite{Beiwenger1976DasTemperatur} It is conjectured that Zn acts as a getter for oxygen, which reduces the internal oxidation of alloy during scale formation, allowing enough time for Al to diffuse to the surface and form $Al_2O_3$, also any $ZnO$ or $Cu_2O$ formed will be reduced by Al oxidation to form  more stable $Al_2O_3$. These studies indicate the third element's strong impact as a facilitator in lowering the required threshold of the primary passivator.

Much of the work described above offered an explanation for the \textit{third element effect} (TEE) at the scale given by molecular species in an oxide slab with no atomic-scale information. Recent works showcase the effects of the Cr--Cr chemical short-range ordering (CSRO) type on the aqueous passivation of binary Fe-Cr solid solution alloys \cite{Liu2018TheAlloys} in 0.1 M $H_2SO_4$ solution. The effect is based on the solid-state oxidation process following a percolation theory-based model proposed by Sieradzki et al. \cite{Sieradzki1986ASteels, Xie2021AAlloys}. This model states that the primary passivator (e.g.,  Cr in Fe-Cr) percolates as $mer$ units of M--O--M until a complete linkage is formed on the surface, resulting in the first monolayer of the passive film. This results in a threshold-type behavior where Cr concentrations $> 13$ at.\% lead to shorter Cr--Cr bond lengths (i.e., better connectivity between Cr atoms), where lesser dissolution of Fe is required in Fe-Cr alloys to collect enough Cr on the surface and form the first monolayer of $Cr_2O_3$ rich passive film. More recent works on quaternary Fe-8Cr-8Al-$x$Ti \cite{Sur2025FactorsTi}, and (FeCoNi)-10Cr-$x$Al  (all at.\%) \cite{Blades2024TuningAlloysb} compositionally complex alloys (CCAs) indicate that the addition of a \textit{third element} such as Al or Ti promotes a clustering type CSRO of the Cr atoms in solid solution. Here, the Cr atoms show shorter bond lengths in the first or second nearest neighbor ($nn$) shells, mimicking the behavior of higher Cr-containing alloys with more enrichment of $Cr_2O_3$ in the alloy passive films. \cite{Sur2025InvestigatingAcid}. The role of Al or Ti here can be termed as the \textit{facilitator} for $Cr_2O_3$ rich passivation. These results provide a novel approach to explain more about the TEE observed in the aforementioned ternary alloys.

In this work, the unresolved aspects of Al-Sn synergy in Cu-Al alloys are revisited and further investigated using a range of arc-melted and solution heat-treated (SHT) ternary Cu-Al-Sn and binary Cu-Al alloys using a suite of electrochemical, spectroelectrochemical, and surface-sensitive techniques such as X-ray photoelectron spectroscopy (XPS)  and scanning transmission electron microscopy (STEM) coupled with energy-dispersive X-ray spectroscopy (EDS). A series of binary and ternary solid solution is largely tested to maintain solid solution alloys and avoid complications of phases with inhomogeneous compositions.  Here, the hypothesis is that the observed Al-Sn synergy is a result TEE type behavior of Sn or its oxides on Al(III), which is the indicated primary passivator. A strong effect of Sn addition is observed in both primary and steady-state passivation behaviors in deaerated {\sulfate}. Using \textit{first-principles} density functional theory (DFT) calculations supported by the cluster expansion formalism and Monte Carlo simulations, the alteration of CSRO in Cu-Al-Sn is also computationally explored. An increase in Al--Al clustering with added Sn is predicted, supporting the $facilitator$ role of Sn with respect to Al following the percolation theory-based model of passivation. Two possible mechanisms are discussed that delineate the role of Sn as a $facilitator$ of $Al_2O_3$ rich passivity of Cu-Al alloys.

\section{Methods}
\subsection{Computational}
\label{subsect:comp}

We explored the effects of local ordering on corrosion behavior previously linked via a percolation model of aqueous passivation \cite{Sieradzki1986ASteels, Xie2021AAlloys, Blades2024TuningAlloysb}. CSRO is the deviation in local atomic correlations from that of the ideal random solid solution, often quantified via the Warren-Cowley parameter \cite{Cowley1950AnAlloys, Cowley1960Short-Solutions}. In multicomponent alloys, Warren-Cowley CSRO parameters $\alpha_{ij}^{p}$ are defined \cite{deFontaine1971TheSystems} for pairs of atom species as
    \begin{equation}
    \alpha_{ij}^{p} =
    \begin{cases}
        \dfrac{P_{ii}^p-\overline{c_i}}{1-\overline{c_i}},  & i = j \\ \\
        1 - \dfrac{P_{ij}^p}{\overline{c_j}}, & i\neq j
    \end{cases}
    \label{eq:WC}
    \end{equation}
where $P_{ij}^p$ is the conditional probability of finding an atom of type $j$ within neighbor shell $p$, given an atom of type $i$ is at the origin, and $\overline{c_i}$ and $\overline{c_j}$ are the average compositions of species $i$ and $j$, respectively.

CSRO was computed using a combination of density functional theory (DFT) \cite{Hohenberg1964InhomogeneousGas, Kohn1965Self-ConsistentEffects}, cluster expansion (CE) \cite{Sanchez_1984, ceder_derivation_1993, deFontaine1994cluster, zunger1994_NATO, kadkhodaei2021cluster,Wu2016review}, and Monte Carlo (MC) simulations. DFT is an \textit{ab initio} method for computing atomic-scale energetics and has been used to compute properties of a wide variety of materials systems \cite{Juergen2008DFT}. The CE is a generalized Ising model that maps atomic arrangements on a lattice to configuration-dependent quantities, most typically the energy of a system. In this study, the CE Hamiltonian was constructed using the Alloy Theoretic Automated Toolkit (ATAT) package \cite{avdw:atat,avdw:atat2,avdw:maps}. To represent multicomponent alloy energetics in the cluster basis, the CE is defined as \cite{avdw:atat,avdw:atat2,avdw:maps,Wrobel_2015_FeCrNi,wolverton1994basis}:

\begin{equation}
    E(\sigma)=\sum_{\alpha} m_{\alpha} J_{\alpha} \overline{\Gamma_{\alpha}}(\sigma)
    \label{E}
\end{equation}

\noindent where $E(\sigma)$ is the energy of configuration $\sigma$,  $\alpha$ is each symmetrically distinct cluster under space group operations of the lattice, $m_\alpha$ are symmetrical multiplicities of each cluster $\alpha$, $J_\alpha$ are effective cluster interactions (ECIs) which are fitted to the DFT data using the structure inversion method \cite{connolly1983SIM}, and $\overline{\Gamma_{\alpha}}(\sigma)$ are the cluster functions averaged over all clusters equivalent to $\alpha$. While there are several choices for cluster function bases (typically orthonormal), the exact form of the basis determines the expressions relating the averaged cluster functions to CSRO parameters. All computed CE quantities and choice of basis functions in this work follow the mathematical definition and derivation in Refs.\ \cite{avdw:atat2, Wrobel_2015_FeCrNi}. The approximate Hamiltonian derived through the CE method can then be used to accelerate the evaluation of energies during MC sampling \cite{connolly1983SIM, kadkhodaei2021cluster, Wu2016review, Sanchez1996book}. ATAT's canonical ensemble MC simulation functions were also used in this work \cite{avdw:emc2}. The CE+MC approach has been extensively used to compute CSRO in numerous binary \cite{wolverton1994basis, Wrobel_2015_FeCrNi, Wolverton2000Short-range-orderStability, Blum2004Mixed-basisAlloys} and multicomponent alloy systems \cite{Fernandez-Caballero2017Short-rangeSystemb,Sobieraj2020ChemicalStudy,Nataraj2021AModels,Chen2024BayesianConstruction}. 

The CE in this work was constructed by calculating Cu-Al-Sn configurations on an underlying FCC lattice. 50 DFT training structures were included in the fit, leading to a cross-validation score of 16 meV/atom. All DFT calculation settings were in accordance with the Open Quantum Materials Database \cite{Saal2013MaterialsOQMD,Kirklin2015TheEnergies} settings, the specifics of which are given in the Supplementary Material. Further details of the CE fitting including the ECI and predicted ground states are also provided in the Supplementary Material. The MC sampling to extract CSRO parameters used a 20$\times$20$\times$20 unit cell with 1,000 MC steps for equilibration, followed by 10,000 for thermodynamic averaging. MC sampling was performed at 3500K; while this temperature is high, overestimation of phase boundary temperatures is common within the CE method \cite{ceder2002}.

\subsection{Experimental}

Alloy specimens with compositions shown in Table \ref{tab:compositions} were synthesized using vacuum arc melting of high purity Cu (electrolytic grade 99.9990 \%),  high purity Sn (99.990 \%), and high purity Al (99.9990 \%) sources into $\approx$ 1 cm diameter button shape. Specimens were encapsulated into Ar purged quartz crucibles to perform a solution heat treatment (SHT) at temperatures mentioned in Table \ref{tab:compositions}, and then quenched in water at room temperature. The SHT sample surface was ground using SiC paper up to a mesh size of \# 1200 and then polished to 1 $\mu$m using polycrystalline diamond solution. Samples were cleaned using isopropanol and deionized water and then dried using $N_{2}$ gas. X-ray diffraction (XRD) phase analysis of the alloy microstructures after SHT was performed using a Bruker D2 Phaser benchtop X-ray diffractometer equipped with Cu K$\alpha$ ($\lambda_{K_{\alpha}}$ = 0.152 nm) X-rays. Alloy microstructures were analyzed using a FEI Quanta 650 scanning electron microscopy (SEM) using the energy dispersive spectrometer (EDS) attachment.

\begin{table}[H]
\centering
\caption{List of all the Cu-Al-Sn alloy SEM-EDS based compositions, their solutionizing heat treatment temperature, and microstructural phases observed from XRD pattern analysis. Here, $\alpha$, $\beta$, and $\beta'$ are FCC (Cu), $Cu_3Al$, and  $Cu_3Al$ martensite structures.}
\label{tab:compositions}
\begin{adjustbox}{width = \textwidth}
\begin{tabular}{cccccc}
\hline
\textbf{Alloy} &
  \textbf{\begin{tabular}[c]{@{}c@{}}Cu\\ (at.\%)\end{tabular}} &
  \textbf{\begin{tabular}[c]{@{}c@{}}Al\\ (at.\%)\end{tabular}} &
  \textbf{\begin{tabular}[c]{@{}c@{}}Sn\\ (at.\%)\end{tabular}} &
  \textbf{\begin{tabular}[c]{@{}c@{}}Solutionizing\\ Temp. ($^\circ$C)\end{tabular}} &
  \textbf{\begin{tabular}[c]{@{}c@{}}Microstructure\\ (XRD)\end{tabular}} \\ \hline
\textbf{Cu-11.5Al}      & 88.5 & 11.5 & -   & 550 & $\alpha$               \\
\textbf{Cu-11.5Al-0.5Sn} & 88   & 11.5 & 0.5 & 550 & $\alpha$               \\
\textbf{Cu-11.5Al-2.5Sn} & 84   & 11.5 & 2.5 & 550 & $\alpha$               \\
\textbf{Cu-16Al}        & 84   & 16   & -   & 550 & $\alpha$               \\
\textbf{Cu-18Al}        & 82   & 18   & -   & 550 & $\alpha$               \\
\textbf{Cu-18Al-0.5Sn}   & 81.5 & 18   & 0.5 & 550 & $\alpha$               \\
\textbf{Cu-23Al}        & 77   & 23   & -   & 750 & $\beta$ + $\alpha$ + $\beta'$  \\
\textbf{Cu-23Al-1.5Sn}   & 76.5 & 23   & 1.5 & 750 & $\beta$ + $\alpha$ + $\beta'$ + $Cu_3Sn$  \\
\textbf{Cu-23Al-3Sn}     & 74   & 23   & 3.0 & 750 & $\beta$ + $\alpha$ + $\beta'$ + $Cu_3Sn$\\ \hline
\end{tabular}
\end{adjustbox}
\end{table}

Electrochemical corrosion investigations were performed using a three-electrode cell consisting of a mercury/mercury sulfate reference electrode (E = +0.640 V vs. standard hydrogen electrode (SHE)), Pt mesh as the counter electrode, and the polished alloy specimens as the working electrode. All potentials herein will be referenced to SHE unless otherwise stated. The working electrode provided a 0.066 $cm^2$ exposed area constrained by a Viton O-ring. Solutions of 0.1 M Na\textsubscript{2}SO\textsubscript{4(aq.)} were prepared using sodium sulfate (Fisher Scientific), and deionized water (Millipore Sigma, 18.2 $M\Omega \cdot cm^2$) was used for all electrochemical experiments. The pH of the solution was adjusted to 4.0 $\pm$ 0.1 using a stock sulfuric acid (96 \%, Sigma Aldrich) solution. $N_2$ gas was bubbled in the solution for 15 min before as well as during the experiments to minimize the effects of dissolved oxygen. A cathodic potentiostatic polarization step was performed at $-$0.56 V for 120 s to reduce or minimize the air-formed (native) oxide. The open circuit potential (OCP) was monitored for 20 s after this step. Anodic LSV experiments were performed from $-$0.1 vs. OCP to 1.2 V vs. SHE at a scan rate of 1 $mV \cdot s^{-1}$. 

Chronoamperometry (CA) tests were performed independently of the LSV experiments after the cathodic reduction step of $-$0.56 V for 120 s. These tests were used to calculate the number of dissolved monolayers ($h$) at an applied potential of $+$0.6 V, a potential selected to be within the passive region for several alloys while recording the current density for 300 s within intervals of 0.01 s.\cite{Xie2021AAlloys,Sur2025FactorsTi,Blades2024TuningAlloysb,Connors2025AnDesign,Anber2025} Following the assumption that primarily consists of \{111\} planes (when FCC matrix) and \{110\} planes (when BCC matrix) were exposed to the solution, $h$ was calculated using the following relation \cite{Xie2021AAlloys}:

    \begin{equation}
    h =  
    \begin{cases}
    
        { \int_{0}^{t} i\, dt} \cdot
	\left[\dfrac{2 \cdot q \cdot ( 1 \cdot X^{bulk}_{Cu} + 3 \cdot X^{bulk}_{Al} +  2 \cdot X^{bulk}_{Sn})}
			{\sqrt{2} \cdot a^{2} }
	\right]^{-1}, \{BCC\} \\ \\

        { \int_{0}^{t} i\, dt} \cdot
	\left[\dfrac{4 \cdot q \cdot ( 1 \cdot X^{bulk}_{Cu} + 3 \cdot X^{bulk}_{Al} +  2 \cdot X^{bulk}_{Sn})}
			{\sqrt{3} \cdot a^{2} }
	\right]^{-1}, \{FCC\}
    
    \end{cases}
    \label{eq:hvalue}
    \end{equation}

where $i$ is the magnitude of current density obtained from CA experiments with total hold time ($t$), $q$ is the electronic charge, ${X^{bulk}_M}$ is the bulk mole fraction of constituent ${M}$ in the alloy, and $\displaystyle{a}$ is the lattice constant of the alloys calculated using XRD pattern analyses. The equation divides the anodic charge density, assumed to result from anodic dissolution, by the charge per monolayer coverage in a unit cell of atoms dissolving congruently. Given the slight change in lattice parameters, $a$, of the alloys, all were assumed to be equal to 0.35 nm. For these calculations, the valence states of Cu, Al, and Sn were assumed to be +2,\cite{Zhou2016} +3, and +2, respectively. Alloys with lower $h$ values indicate rapid, efficient re-passivation such that there is less anodic dissolution required by the alloy for accumulating enough passivator constituents (such as Al) on the surface to form a stable passive film.\cite{Xie2021AAlloys}  

In a third procedure, passive film CA growth experiments were performed for 10 ks by directly stepping to $+0.6$ V (within the passive region for a majority of the alloys) from the cathodic minimization pre-treatment step described above. Potentiostatic electrochemical impedance spectroscopy (EIS) was performed after the CA hold of 10 ks at $+0.6$ V. A sinusoidal signal of 20 mV (RMS) was applied at a $+0.6$ V potential across a frequency range of 100 kHz to 5 mHz at 6 points per decade of frequency. During the hold, in addition to current density, the imaginary component of impedance (\zimag) was also recorded for every 200 ms using a sinusoidal signal of 20 mV (RMS) and a single frequency of 5 Hz to evaluate qualitative trends in the passive film thickness using relationships developed elsewhere. \cite{Benoit2016ComparisonMetals, Hirschorn2010Constant-Phase-ElementApplications, Hirschorn2010Constant-Phase-ElementFilms, Sur2025InvestigatingAcid} 
Finally, long-term corrosion resistance was probed via passive film impedance by exposing Cu-23Al-3Sn, Cu-23Al, high purity Cu samples at OCP for 28 days. A Hirox RH-8800 Light Microscope was used to obtain optical images of the corroded surface. 

Specific elemental dissolution current densities ($i _M$) of Cu, Al, and Sn were tracked for Cu-23Al and Cu-23Al-3Sn during LSV in the anodic direction (hereby termed as AESEC-LSV), as well as CA at $+0.6 $ V for 2.5 ks (hereby termed as AESEC-CA), both in deaerated {\sulfate} using an AESEC apparatus described elsewhere.\cite{Ogle2019AtomicPassivation} Both AESEC-LSV and AESEC-CA experiments were performed after the cathodic potentiostatic polarization step at $-0.16$ V for 300 s to minimize the presence of air-formed (native) oxide. For AESEC analysis, an inductively coupled plasma atomic emission spectrometer (ICP-AES, Horiba France) equipped with a polychromator with a focal length of 0.5 m and a monochromator with a focal length of 1.0 m was used to detect elements dissolved in the electrolyte during electrochemical experiments. The elemental Sn signal of wavelength 190 nm was obtained by the monochromator to improve its resolution. The emission intensities ($I_\lambda$) at characteristic wavelengths ($\lambda$) measured by the spectrometer for each element M was converted to concentration ($C_M$) in $nmol\cdot cm^{-3}$ using ICP-AES calibration method of using blank electrolyte, and standard electrolytes containing $0.2$ ppm and $0.4$ ppm of the elements to obtain a calibration curve. 

\begin{equation}
     C_M = (I_\lambda - I^\circ _ \lambda) / k_{\lambda,M}
\end{equation}
where $I^\circ_\lambda$ and $k_{\lambda,M}$ are the background intensity and the
sensitivity factor for element M, respectively, for a given wavelength. The concentration was, in turn, converted into dissolution current density ($i_M$) :

\begin{equation}
    i_M = n_M \cdot F\cdot f \cdot C_M / A
    \label{eq:dissolution rate}
\end{equation}

\noindent Here, $n_M$ is the valence state of the dissolved metal M mentioned above for Cu, Al and Sn, $F$ is Faraday's constant, $f$ is the electrolyte flow rate of approximately $0.02$ $cm^3\cdot s^{-1}$ in the electrochemical flow cell, and $A$ is the flow cell exposed surface area of 1 $cm^2$. Total dissolution current density $i_{sum}$ was defined as $\sum i_M$. Convoluted electrical current densities ($i^*_e$) were obtained from actual electrical current density ($i_e$) considering the electrolyte flow hydrodynamics between the electrochemical cell and ICP-AES is elaborated elsewhere.\cite{Ogle2019AtomicPassivation} Assuming Cu undergoes congruent dissolution in these alloys, hypothetical congruent dissolution current density ($i_M ^{cg}$) for Al and Sn were obtained using the relation:

\begin{equation}
     i_M ^{cg} = \dfrac{n_M\cdot X_M}{n_{Cu} \cdot X_{Cu}} \cdot i_{Cu} 
    \label{eq:cg}
\end{equation}

\noindent Here, $X_M$ is the mole fraction of element M, and $M = \{ Al, \ Sn \}$. The difference between the congruent and actual dissolution rate indicates the element's surface enrichment rate. The differences between the area under the curves $i _M ^{cg}$ and $i _M$ for a given interval of potential or time would indicate the elemental enrichment (${\Theta}_M$, in $C\cdot cm^{-2}$) at the surface as defined below.  
\begin{equation}
      \Theta_M =  \int_a^b (i^{cg}_M \space - i_M ) \ dt  
    \label{eq: theta}
\end{equation}
Here, $a$ and  $b$  are the lower and upper limits of time intervals, which may also correspond to potential intervals in an AESEC-LSV experiment as the potential sweep is linear with time. For an element M,  ${\Theta}_M > 0$ indicates surface enrichment, while ${\Theta}_M < 0$ signifies surface depletion.

\begin{figure}[H]
    \centering
    \includegraphics[width=0.7\linewidth]{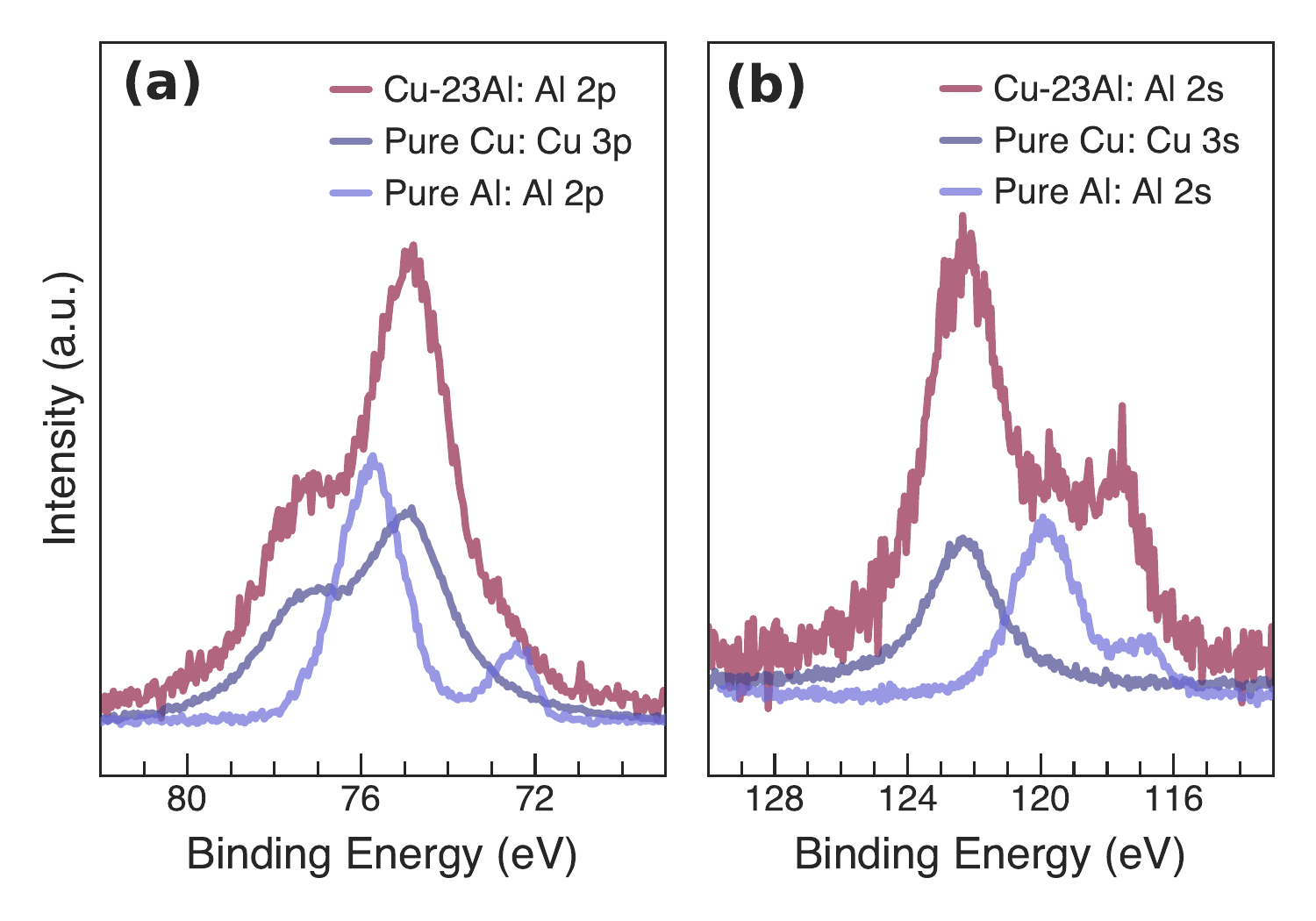}
    \caption{XPS core-shell spectral data showing overlaps in (a) Al 2p and (b) Al 2s spectral regions of a Cu-23Al (at.\%) alloy compared to the core-shell spectra of the same regions obtained from the high purity Cu and the high purity Al sample.} 
    \label{fig:xpsoverlap}
\end{figure}

Chemical compositions and sputter depth profiling of oxide passive films after CA experiments at +0.6 V and 300 s were analyzed using X-ray photoelectron spectroscopy (XPS). Specimens were transferred within 5 min in $N_2$ gas filled glove bags. Cationic species in the electrochemically grown passive film were analyzed using a PHI VersaProbe-III XPS Analyzer, calibrated with an Au standard to the 4f\textsubscript{7/2} core level at 84.00 eV binding energy. C 1s (284.80 eV) was used as a reference to shift spectra for any correction due to surface charging. Core-shell spectra of O 1s, Cu $2p_{3/2}$, Sn $3d_{5/2}$, Al 2s, and Auger electron spectra of Cu $L_3M_{45}M_{45}$ were deconvoluted to obtain cationic compositions as done previously.\cite{Glover2022InterrogatingApproach} Al 2s region was chosen due to its smaller overlap with Cu 3s compared to that of Al 2p with Cu 3p, see Figure \ref{fig:xpsoverlap}. The overlap of Cu 3s was constrained by fixing the area of its peak proportional to the total peak areas of the Cu $L_3M_{45}M_{45}$ Auger spectra by multiplying the ratio of their relative instrument sensitivity factors provided by PHI (see Table S1 of the supplementary material). Voigt and Doniach-Sunjic convoluted Gaussian peaks were used to fit cationic and metallic peaks, respectively, following the reported peak positions, relative peak amplitudes, and number of peaks for Cu LMM \cite{Biesinger2017}, Al 2s \cite{Gall2013BoronCells} and Sn $3d_{5/2}$ \cite{Jie1990XPSSurface, Venugopal1997X-RaySolution}.

The passive film grown on the Cu-23Al-3Sn sample for XPS studies was also examined via STEM-EDS using a JEOL F200 200 kV TEM. The TEM lamellae were prepared using a Focused Ion Beam (FIB) with a HELIOS G5 UC dual-beam system. The samples underwent a standard \textit{in-situ} FIB lift-out procedure using 30 kV Ga\textsuperscript{+} ions and were subsequently thinned to electron transparency at 2 kV.

\section{Results}

\subsection{Simulated chemical short-range order in Cu-Al-Sn}
\label{subsec:simSRO}

\begin{figure}[H]
    \centering
    \includegraphics[width=0.75\linewidth]{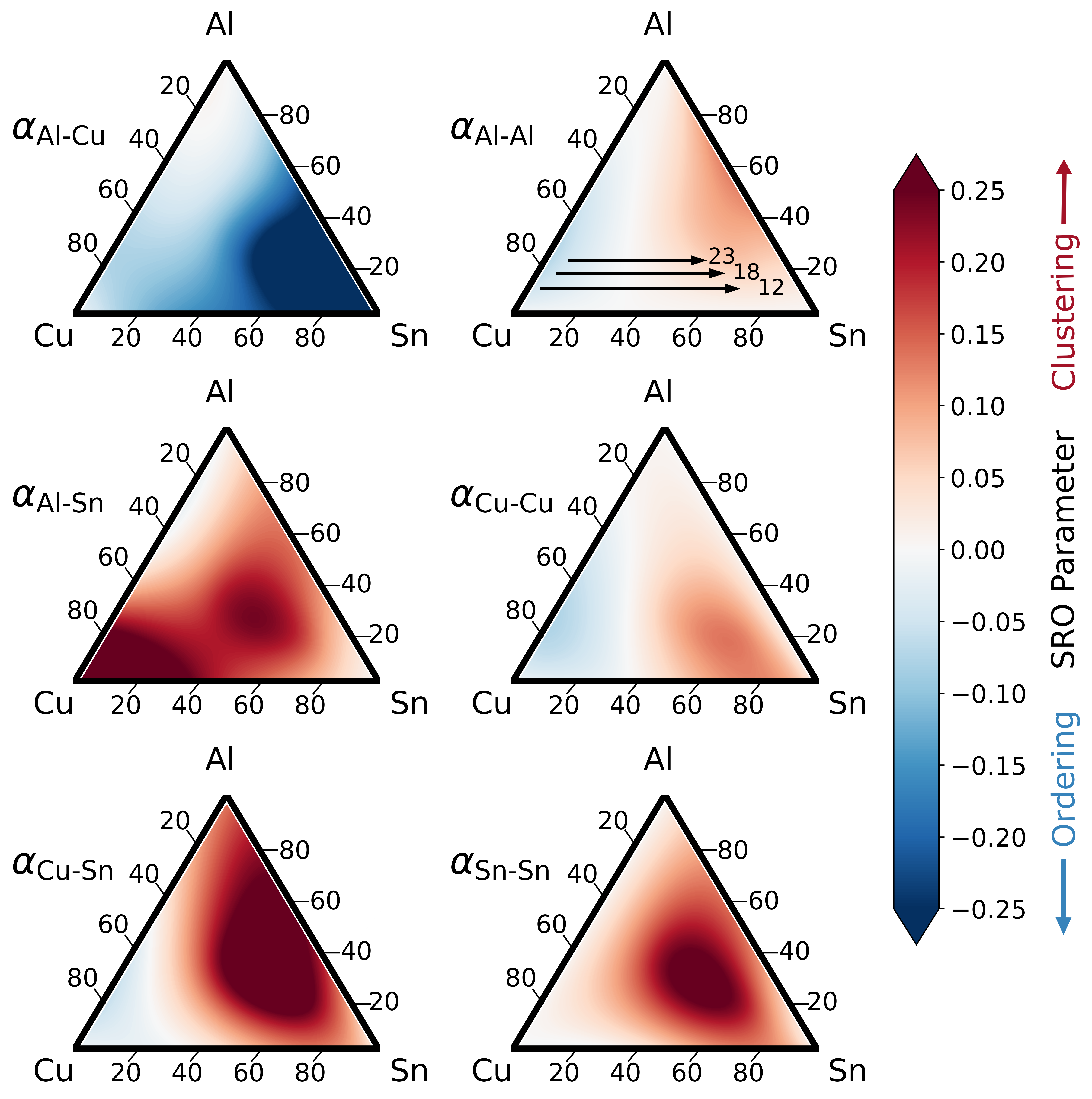}
    \caption{Ternary diagrams of Al-Cu, Al-Al, Al-Sn, Cu-Cu, Cu-Sn, Sn-Sn CSRO obtained from first principles cluster expansion and Monte Carlo simulations in FCC Cu-Al-Sn alloys. The adjacent color bar indicates the Warren-Cowley $\alpha_{ij}$ CSRO parameter. Experimentally studied compositions Cu-0.12Al-$y$Sn, Cu-0.18Al-$y$Sn, and Cu-0.22Al-$y$Sn (all in mole fraction) are indicated as horizontal arrows on the Al--Al ternary diagram.}
    \label{fig:csromap}
\end{figure}

Figure \ref{fig:csromap} demonstrates the CE-derived Warren-Cowley $\alpha_{ij}$ values of all possible like and unlike elemental pairs as a function of ternary Cu-Al-Sn composition for the first neighbor shell. The primary focus is on the Al--Al pair, as it is found to be the primary passivating constituent in the Cu-Al-Sn alloys and critical to passivation and corrosion resistance based on previous studies as well as work herein.\cite{Sur2025FactorsTi, Sur2025InvestigatingAcid, Blades2024TuningAlloysb, Liu2018TheAlloys, Anber2025} Along the Cu--Al edge (i.e., lean Sn) of the ternary, $\alpha_{\text{Al--Al}}$ always shows small negative values, indicating a slight ordering behavior. With increasing Sn content for fixed Al content (see arrow in Figure \ref{fig:csromap}), $\alpha_{\text{Al--Al}}$ increases, changing sign from negative to positive, indicating a change in CSRO tendency from ordering to clustering type. Among other pairs, Cu--Cu also shows a small tendency to order that changes to a clustering tendency with increasing Sn content. Sn--Sn pairs show a tendency to cluster at all compositions. Further, Al--Cu shows a strong tendency to order, indicating that Al and Cu preferentially bond with each other. This behavior aligns with observations of the precipitation behavior of Al-rich Al-Cu alloys.\cite{Sha2011NanostructureProcesses} For the experimentally studied compositions in this work, Al--Al ordering tendency decreases with Sn addition up to 3 at.\%  suggesting more availability of Al and Sn (clustering of Sn--Sn) for aqueous passivation processes. 

\subsection{Alloy microstructure}
Cu-Al-Sn alloys with 18 at.\% Al or lower reflected a single phase FCC microstructure as listed in Table \ref{tab:compositions} from X-ray diffraction phase analyses shown in Figure S2, which match the expected CALPHAD phase diagram shown in Figure S1. Optical imaging revealed an average grain size of approximately 400 microns for all the alloys. This is expected as all the alloys were prepared following the same heat treatment procedure. However, Cu-23Al alloys with and without Sn added did not show a single-phase microstructure as predicted by CALPHAD, see Figures S1-S4 in the supplementary material, where, other than the expected phase of $Cu_3Al$ structure, $Cu_3Al$ martensite, FCC Cu, as well as some $Cu_3Sn$ phases were observed. These multi-phase alloys were primarily homogeneous in composition. Although such a microstructure with partitioned passivators into phases might degrade the passivity, the Cu-23Al-Sn alloys showed more efficient passivation and higher film impedance (discussed in Section \ref{sec: impedance} than Cu-23Al, both showing a martensitic microstructure.\cite{Inman2024VariationCorrosion} This suggests that between these alloys, the compositional effects from Sn addition on alloy passivity are more impactful than those of the changes in microstructural factors.

\subsection{Evaluations of corrosion electrochemistry}
\subsubsection{Anodic linear sweep voltammetry to assess E-I behavior}
High purity Cu and Sn metals, and binary Cu-Sn alloys exhibited active corrosion across all anodic potentials under a charge transfer controlled E-I behavior from  $0.2$ V to $0.35$ V, positive to corrosion potential ($E_{corr}$) in deaerated {\sulfate} as shown in Figure \ref{fig:lsv}a. Further, a mass transfer limited region (with current densities greater than 10 $mA \cdot cm^{-2}$)  at large positive overpotentials. Compared to the LSV behavior of high purity Cu, high purity Sn showed greater anodic current densities and a more negative $E_{corr}$ $\approx -0.35$ V. This difference in the LSV behaviors of high purity Cu and high purity Sn helped explain the behavior for binary Cu-Sn alloys where Cu-0.5Sn (at.\%) and Cu-2.5Sn (at.\%) alloys showed greater anodic current densities than high purity Cu. On the contrary, high purity Al exhibited passivation with a charge transfer limited region regulated by Al(III) cation-rich passive film (cation and anion transport through hydrated $Al_2O_3$) at a lower potential of $\approx -1$ V and an {\ipass} of $\approx 0.02$ $mA \cdot cm^{-2}$.

Figure \ref{fig:lsv}a also documents the anodic LSV behaviors of Cu-Al-Sn alloys in deaerated {\sulfate}. Only select alloys with Al content higher than 18 at.\% showed passive current densities (\ipass) lower than 1 $mA \cdot cm^{-2}$.  Compared with high purity Cu's behavior, the 11.5 at.\% Al added Cu alloy showed an insignificant change in the anodic behavior with no active-passive transition. However, higher Al additions of 16, 18, and 23 (at.\%) to Cu resulted in a distinct active-passive transition where {\icrit} decreased by an order of magnitude with increasing Al content. Interestingly, additions of Sn in the ternary alloy introduced an active-passive transition feature and lowered the anodic current densities depending upon the Al and Sn contents. 
Compared to the LSV behavior of Cu-11.5Al, Cu-11.5Al-0.5Sn shows a small change in current density at $\approx +0.9$ V. Moreover,  Cu-11.5Al-2.5Sn (at.\%) shifts this potential to $\approx +0.6$ V while resulting in a smaller {\icrit} of $\approx 40$ $mA \cdot cm^{-2}$. This benefit of lowering of {\icrit} and {\ipass} was also observed when considering Sn addition in Cu-18Al and Cu-23Al alloys, such that the Cu-23Al-3Sn alloy exhibited an {\icrit} value of $\approx 0.07$  $mA \cdot cm^{-2}$ and {\ipass} value of $\approx 0.015$  $mA \cdot cm^{-2}$, approaching stainless steel like {\ipass}, which are $\approx 3$ and $\approx 5$ times smaller than that shown by Cu-23Al alloy, respectively as shown in Figure \ref{fig:lsv}a. Further, this alloy, when compared to the anodic LSV behavior of high purity Al, showed comparable yet slightly smaller {\ipass} values.

\subsubsection{Number of dissolved monolayers (\textit{h}) required to achieve passivation}
\begin{figure}[H]
    \centering
    \includegraphics[width=0.75\linewidth]{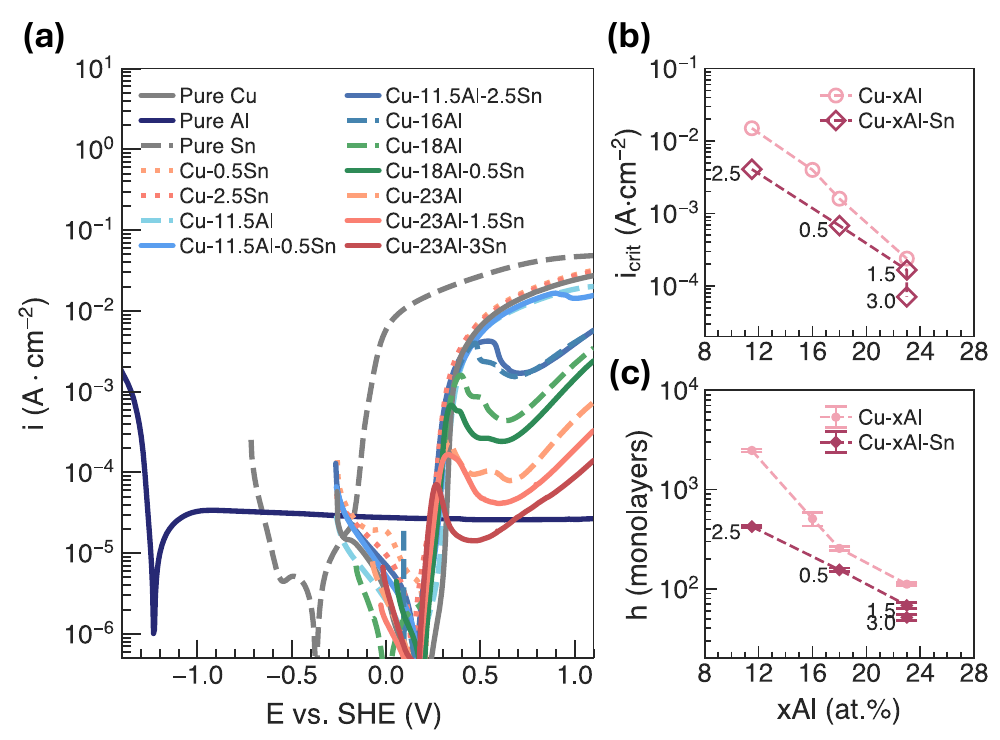}
\caption{(a) Anodic linear sweep voltammetry $log(i)$ vs. $E$ curves of high purity Cu, high purity Al, high purity Sn, Cu-0.5Sn, Cu-2.5Sn, Cu-23Al-3Sn  alloys, and Cu-Al-Sn alloys (all at.\%) after cathodically reducing the air-formed oxides. (b) Critical current density ($i_{crit}$) values extracted from the curves plotted against Al content (at.\%) in the alloy. (c) No. of dissolved alloy monolayers ($h$) calculated for Cu-$x$Al-Sn alloys from CA experiments at $+0.6$ V. All experiments were performed in deaerated \sulfate.
}
\label{fig:lsv}
\end{figure}

The number of dissolved monolayers or $h$ values of Cu-Al-Sn alloys plotted against the Al alloying content obtained using single-step CA experiments at $+0.6$ V in {\sulfate} and Equation \ref{eq:hvalue} are shown in Figure \ref{fig:lsv}c. Within Cu-Al binary alloys, increasing the Al content from 11.5 at.\% to 23 at.\% brought about a considerable decrease in $h$ values from $\approx 7000$ to $\approx 150$ monolayers, indicating the impact and supporting the notion of a role of Al as a primary passivator when added in solid solution with Cu. The addition of Sn to the binary Cu-Al alloys showed a significant and consistent further decrease in $h$ values across the range of low to high Al compositions. Cu-11.5Al-0.5Sn showed $\approx 7$ times smaller $h$ values of $\approx 1050$ monolayers, compared to Cu-11.5Al. For Cu-18Al-0.5Sn, the alloys showed an $h$ value of $\approx 250$, which is 2 times smaller than that of Cu-18Al. Cu-23Al-1.5Sn and Cu-23Al-3Sn both exhibited low $h$ values of $\approx 90$ and $\approx 70$ monolayers, respectively, $\approx 1.5$ and $\approx 2$ times smaller than that of Cu-23Al. This suggests that Cu-$x$Al-Sn alloys form passive films more efficiently by requiring fewer monolayers of metal dissolution to reach the conditions where passivation could be obtained for an alloy containing such an Al content \cite{Xie2021AAlloys, Blades2024TuningAlloysb}.

\subsubsection{\textit{In-operando} tracking of elemental dissolution rates during passivation }\label{sec:aesec}

\begin{figure}[H]
    \centering
    \includegraphics[width=0.9\linewidth]{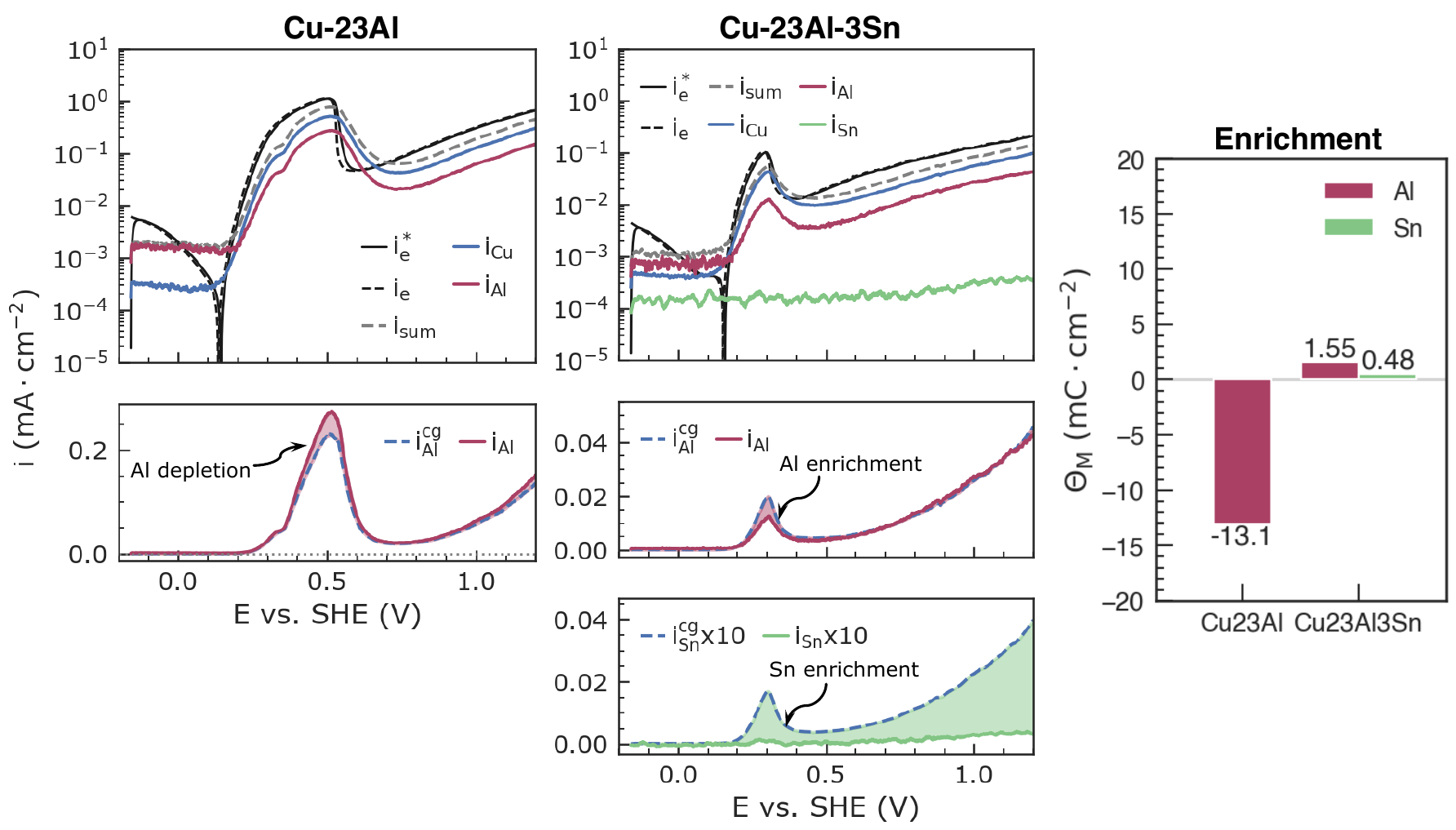}
    \caption{Actual ($i_e$) and convoluted ($i^*_e$) electrical current densities, spectral dissolution ($i_{M}$),  and congruent  ($i^{cg}_{M}$) dissolution current density profiles of Cu, and Al, from Cu-23Al alloy and Cu, Al, and Sn from Cu-23Al-3Sn alloy, both obtained using AESEC-LSV experiments in deaerated {\sulfate}. 
   \%(c) The same profiles for Cu-23Al-3Sn were obtained from AESEC-CA experiment at +0.6 V for 3 ks.
    `Snx10' in the legend means the values shown are the true Al dissolution rate multiplied by a factor of 10. A boxcar averaging (N = 20) method was used for the Sn curves to improve the signal-to-noise ratio. Total surface enrichment ($\Theta_M$) calculated for Al and Sn from AESEC-LSV experiments within the potential range of $E_{corr}$ to +0.6 V.}
    \label{fig:AESEC-LSV}
\end{figure}

Figure \ref{fig:AESEC-LSV} reports the Al, Sn, and Cu dissolution current densities in solution of Cu-23Al and Cu-23Al-3Sn alloys during AESEC-LSV experiments in deaerated {\sulfate}. The electrical current density profiles ($i_e$) are similar to those observed in the LSV experiments described in Figure \ref{fig:lsv}a with smaller {\icrit} and {\ipass} observed for Cu-23Al-3Sn alloy in comparison to Cu-23Al.In this section, $i^{cg}_{Al}$ and $i^{cg}_{Sn}$ were calculated using Eq. \ref{eq:cg} as individual elemental current densities for the dissolved species, assuming congruent dissolution.
Both alloys showed some dissolution of pre-existing air-formed oxides in the cathodic region of the LSV. In the anodic region, Cu-23Al-3Sn showed significantly lower $i_M$ values for all Cu, Al, and Sn compared to Cu-23Al alloy, with Cu dissolved at $\approx 10$ times and Al at $\approx 25$ times lower rates than in Cu-23Al. Specifically in the active-passive transition peak region, for Cu-23Al, the $i_{Al}$ profile was larger than its $i^{cg}_{Al}$ profile, indicating Al surface depletion (negative enrichment). In contrast, Cu-23Al-3Sn exhibited a lower $i_{Al}$ profile than its $i^{cg}_{Al}$ profile, suggesting enrichment of Al on the surface. Sn enrichment was also observed for Cu-23AlSn with $i_{Sn}$ being much lower than $i^{cg}_{Sn}$ in the entire anodic region. Further it should be noted that Cu-23Al showed a potential range of passive film instability of approx. 200 mV, where contributions of cationic dissolution from the passive film are observed (i.e., $i_{sum} > i^*_e$) at potentials just higher than the anodic peak potential.\cite{Ogle2019AtomicPassivation} For Cu-23Al-Sn, this potential range was halved to approx. 100 mV, indicating a benefit in passive film stabilization from Sn presence.

The accumulated total surface enrichment of Al ($\Theta_{Al}$) and Sn ($\Theta_{Sn}$) was calculated for Cu-23Al and Cu-23Al-3Sn in regions of interest using Eq. \ref{eq: theta}, shown in Figure \ref{fig:AESEC-LSV}. For Cu-23Al, ($\Theta_{Al}$) during the active-passive transition peak in AESEC-LSV was $\approx -13.1$, indicating depletion of Al. In contrast, for Cu-23Al-3Sn, it was $\approx +1.55$, and during the period of 20--1000 s of the AESEC-CA experiment, it was $\approx 2.69$ (see Figure S6 in the supplementary material for AESEC-CA results). This indicates that Sn not only reduced Al surface depletion but also contributed to its positive enrichment during the passive film formation process. $\Theta_{Sn}$ was found to be $\approx 0.48$ and $\approx 0.77$ in the above-described regions of interest AESEC-LSV and AESEC-CA, respectively. It should be noted that $\Theta_{Sn}$ is smaller than $\Theta_{Al}$ for both situations.

\subsubsection{Passive film formation investigated through electrochemical impedance}\label{sec: impedance}
\begin{figure}[H]
    \centering
    \includegraphics[width=\linewidth]{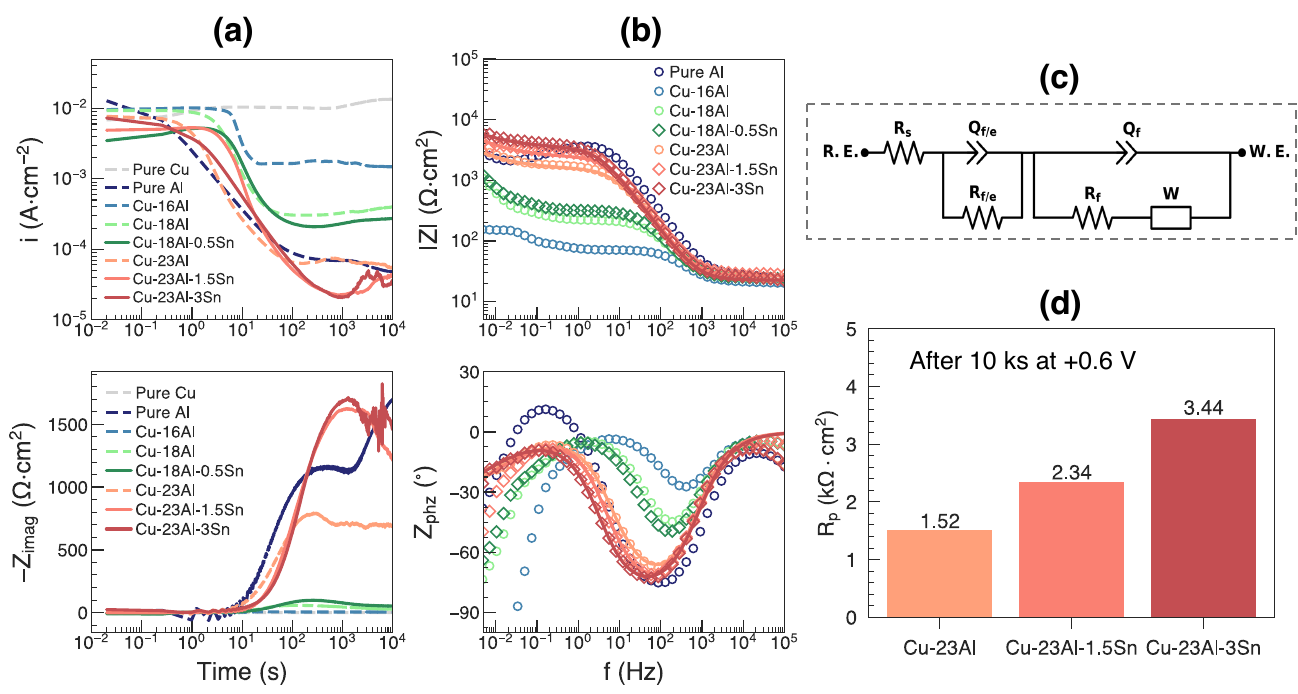}
    \caption{(a) Current density ($i$) and {\zimag} impedance profiles for all the Cu-Al-Sn alloys, high purity Cu and high purity Al metals in deaerated obtained from the single-step CA experiments performed at $+0.6$ V for 10 ks, and (b) Bode impedance modulus ($|Z|$) and phase ($Z_{phz}$) plots obtained from potentiostatic EIS experiments performed at $+0.6$ V after the 10 ks hold in deaerated {\sulfate}. The solid lines indicate spectra obtained after fitting with the ECM.(c) Schematic of the ECM used. Here, $R_s$, $R_{f/e}$ and $Q_{f/e}$, $R_f$ and $Q_f$, and $W$ represent the solution resistance, film/electrolyte interface resistance and constant phase element (CPE), and the film resistance and CPE, and Warburg diffusional impedance element, respectively. (d) $R_p$ values obtained for Cu-23Al, Cu-23Al-1.5Sn, and Cu-23Al-3Sn alloy samples. Here, $R_p = R_{f/e} + R_f$.}
    \label{fig:sfeisandeis-cualsn}
\end{figure}

Figure \ref{fig:sfeisandeis-cualsn}a illustrates the current density ($i$) and {\zimag} values of select Cu-Al-Sn alloys, high purity Cu, and high purity Al obtained using single-step CA hold at $+0.6$ V for 10 ks in deaerated {\sulfate}. All data were fit to the equivalent electrical circuit model (ECM) shown in Figure \ref{fig:sfeisandeis-cualsn}c. Recall that {\zimag} at select frequencies dominated by oxide properties is directly proportional to oxide thickness if assumed to behave like a parallel plate capacitor.\cite{Hirschorn2010Constant-Phase-ElementApplications, Hirschorn2010Constant-Phase-ElementFilms} For all samples, the large $i$ and extremely small {\zimag} values at short times (i.e., $0.01 - 1$ s) may indicate a lack of passive film formed. At intermediate times, i.e., between $1-100$ s, alloys containing 23 at.\% Al and high purity Al demonstrated a significant increase in {\zimag} and a parallel decrease in $i$, suggesting protective passive film formation.\cite{Sur2025InvestigatingAcid} Other alloys containing less than 23 at.\% Al showed a small rise in {\zimag} and small decay in $i$ remaining $> 0.01$  $mA \cdot cm^{-2}$ over this period. High purity Cu, unlike high purity Al, demonstrated no indication of passivity within a region of charge transfer control before likely mass transport dissolution with $i$ values remaining relatively constant at 10 $mA \cdot cm^{-2}$ throughout the experiment. Interestingly, Sn-added to Cu-23Al alloys showed a 5 times smaller current decay and a larger {\zimag} than Cu-23Al alloys, suggesting the formation of a relatively thicker and more protective passive film provoked by Al and Sn alloying in Cu. The overall difference between the behaviors of Cu-23Al-1.5Sn and Cu-23Al-3Sn was minimal, where the latter showed slightly smaller $i$ and larger {\zimag} values throughout the hold of 10 ks. These results suggest that minor Sn addition to high Al-containing Cu alloys is beneficial in significantly reducing Cu corrosion over time.

Figure \ref{fig:sfeisandeis-cualsn}b exhibits the full spectrum EIS Bode impedance modulus ($|Z|$) and phase ($Z_{phz}$) obtained for the select alloys obtained at $+0.6$ V after 10 ks of single-step CA hold at $+0.6$ V.In this paragraph, $|Z|$ at 5 mHz is used for the sake of the discussion. High purity Al and Cu-23Al based alloys showed $|Z|$ values at 5 mHz of $\approx 5$ k$\Omega \cdot cm^2$, which are $\approx 25$ times and $\approx 5$ times larger than that of Cu-16Al, and Cu-18Al based alloys, respectively. High purity Cu, unlike high purity Al, showed the lowest $|Z|$ value of $\approx 0.02$ k$\Omega \cdot cm^2$, suggesting no protective passive film was present on its surface. A general increasing trend can be observed within $|Z|$ values against the Al content in the alloys. Further, 1.5 at.\% of Sn addition to Cu-18Al resulted in a slightly larger $|Z|$ value of $\approx 1.25$ k$\Omega \cdot cm^2$. A significant increase in $|Z|$ values was observed between the Cu-23Al-3Sn and Cu-23Al-1.5Sn alloys, which showed values of $\approx 4 $ k$\Omega \cdot cm^2$ and $\approx 6 $ k$\Omega \cdot cm^2$, respectively, compared to $\approx 2$ k$\Omega \cdot cm^2$ the Cu-23Al alloy. Figure \ref{fig:sfeisandeis-cualsn}c shows the equivalent circuit model (ECM) \cite{Jakupi2011TheSolution} used to obtain polarization resistance ($R_p$) values of the passive films of Cu-23Al, Cu-23Al-1.5Sn, and Cu-23Al-3Sn shown in Figure \ref{fig:sfeisandeis-cualsn}d. This ECM is reportedly used for fitting passive film EIS spectra.\cite{Sur2025FactorsTi, Gerard2020AqueousFilm, Inman2023CorrosionPassivation, Inman2024EffectCCAs, Lutton2023PassivationSolutions} The $R_p$ of these samples shows the same trend observed from their $|Z|$ values at 5 mHz, demonstrating the benefit of alloying with Sn. Values of the individual ECM elements can be found in Table \ref{tab:ecm-values-cualsn}. Overall, electrochemical analysis suggests a synergistic benefit between Al and Sn in achieving improved stainless behavior.

\begin{table}[H]
\centering
\caption{Values of ECM elements for Cu-23Al, Cu-23Al-1.5Sn, and Cu-23Al-3Sn alloys obtained from fitting the EIS data shown in Figure \ref{fig:sfeisandeis-cualsn}b.}
\label{tab:ecm-values-cualsn}
\begin{adjustbox}{width = \textwidth}
\begin{tabular}{@{}cccccccccc@{}}
\toprule
\textbf{Sample} &
  \textbf{\begin{tabular}[c]{@{}c@{}}$R_s$\\ ($\Omega \cdot cm^{2}$)\end{tabular}} &
  \textbf{\begin{tabular}[c]{@{}c@{}}$Y_{f/e}$\\ ($\mu S\cdot cm^{-2}\cdot s^{\alpha}$)\end{tabular}} &
  \textbf{$\alpha_{f/e}$} &
  \textbf{\begin{tabular}[c]{@{}c@{}}$R_{f/e}$\\ (k$\Omega \cdot cm^{2}$)\end{tabular}} &
  \textbf{\begin{tabular}[c]{@{}c@{}}$Y_{f}$\\ ($\mu S\cdot cm^{-2}\cdot s^{\alpha}$)\end{tabular}} &
  \textbf{$\alpha_{f}$} &
  \textbf{\begin{tabular}[c]{@{}c@{}}$R_{f}$\\ (k$\Omega \cdot cm^{2}$)\end{tabular}} &
  \textbf{\begin{tabular}[c]{@{}c@{}}$W_o$\\ (k$\Omega \cdot cm^{2} \cdot s^{-1/2}$)\end{tabular}} &
  \textbf{\begin{tabular}[c]{@{}c@{}}$R_{p}$\\ (k$\Omega \cdot cm^{2}$)\end{tabular}} \\ \midrule
\textbf{Cu-23Al}         & 22.3  &  37.5  & 0.94 &  0.87  & 46.6 & 0.85 & 0.65 & 0.19 &  1.52  \\
\textbf{Cu-23Al-1.5Sn}   & 26.3  &  39.1 & 1.0 & 1.0 & 26.2 & 0.89 & 1.34 & 0.31 & 2.34\\
\textbf{Cu-23Al-3Sn}     & 22.4  &  44.4 & 0.98 & 0.98 & 34.1 & 0.87 & 2.46 & 0.35 & 3.44 \\ \bottomrule
\end{tabular}
\end{adjustbox}
\end{table}

\begin{figure}[H]
    \centering
    \includegraphics[width=\linewidth]{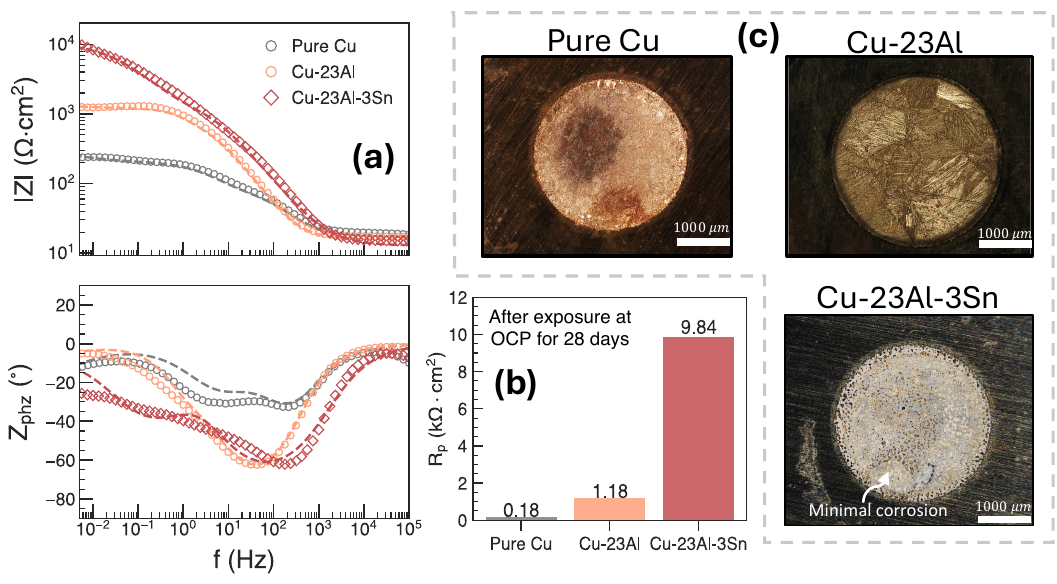}
    \caption{(a) Bode impedance modulus (\zmod) and phase ($Z_{phz}$) plots along with their fit data (dashed lines) and (b) $R_p$ values obtained from  ECM fitting of the EIS data performed on high-purity Cu, Cu–23Al, and Cu–23Al–3Sn (at.\%) alloys at OCP after 28 days of exposure at OCP in deaerated {\sulfate}. (c) Post exposure optical images of the exposed region on the sample surface. }
    \label{fig:eis-ocp}
\end{figure}

Figure \ref{fig:eis-ocp}a demonstrates the impedance behavior at OCP of high purity Cu, Cu-23Al, and Cu-23Al-3Sn alloys after 28 days of exposure in deaerated {\sulfate}. At 5 mHz, Cu-23Al-3Sn exhibited an approximately tenfold higher {\zmod} value of $10$ k$\Omega \cdot cm^2$ compared to $1.5$ k$\Omega \cdot cm^2$ for Cu-23Al, which in turn, displayed an approximately tenfold higher {\zmod} value compared to $0.25$ k$\Omega \cdot cm^2$ for high-purity Cu. Using the ECM shown in Figure \ref{fig:sfeisandeis-cualsn}c, impedance data were fit to obtain $R_p$ values shown in Figure \ref{fig:eis-ocp}b. These values obtained from ECM fitting closely match with {\zmod} values at 5 mHz, as well as the trend between the behaviors. The values of the ECM circuit elements can be found in Table S1 of the supplementary material. Post-exposure surface imaging of the Cu-23Al-3Sn sample showed lesser loss of material compared to high purity Cu and Cu-23Al where the corrosion depth was higher, as shown in Figure \ref{fig:eis-ocp}c. These results indicate that the surface of the Cu-23Al-3Sn sample is significantly more protective over time than that on Cu-23Al, highlighting the consistent beneficial effect of small Sn addition even after 28 days of exposure.

\subsection{Oxide passive film composition}
\subsubsection{XPS spectral deconvolution and peak fitting}

\begin{figure}[H]
    \centering
    \includegraphics[width=\linewidth]{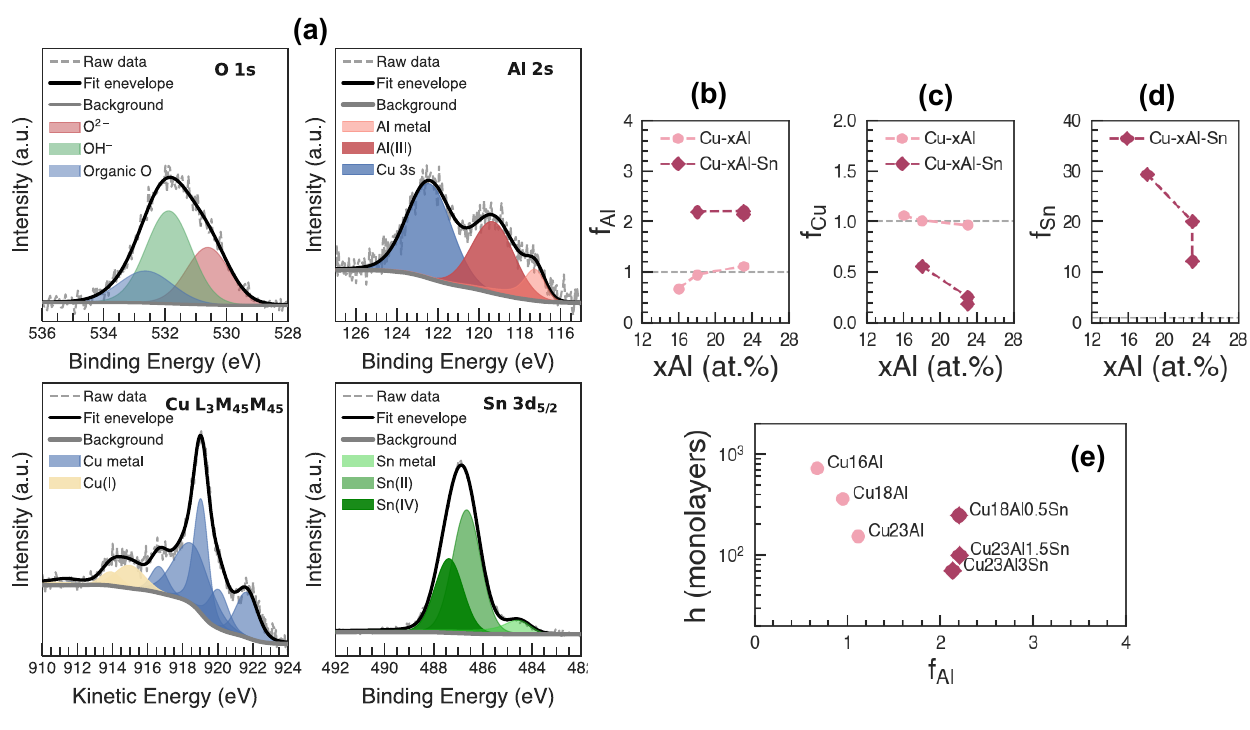}
    \caption{ (a) XPS spectral deconvolution peak fits of O 1s, Al 2s, and Sn $3d_{5/2}$ core-shell spectra and Cu $L_3M_{45}M_{45}$ Auger spectra. (b-d) cationic enrichment factor ($f$) of Al, Cu, and Sn plotted against Al content calculated using cation fractions obtained from XPS peak fitting of the alloy passive films electrochemically grown via CA experiments at $+0.6$ V for 300 s in deaerated {\sulfate}. (e) Number of dissolved monolayers ($h$) values also obtained from the same CA experiments plotted vs. $f_{Al}$.
    }
    \label{fig:xps}
\end{figure}

Cationic compositions of Cu(II, I), Al(III), and Sn(IV, II)  species of the electrochemically grown passive films of select Cu-Al-Sn alloys were determined after single-step CA experiments at $+0.6$ V for 300 s via peak fitting of their XPS spectral regions. Fit spectra for films grown on each alloy in deaerated {\sulfate} identified Cu(I), Cu(II), Al(III), Sn(II), and Sn(IV) cations as well as $OH^-$ and $O^{2-}$ species. Metallic Cu, Al, and Sn species were observed from the alloy beneath the passive films, which suggests the film was a few nanometers in thickness.\cite{Strohmeier1990AnAlloys} Representative spectral deconvolution peak fits from the Cu-23Al-3Sn sample are shown in Figure \ref{fig:xps}a. The O 1s is consistent with $O^{2-}$ and $OH^{-}$, typical of oxy/hydroxides in passive films. Cu was consistent with Cu(I), with some presence of Cu(II) oxides and hydroxides observed. Sn is detected as Sn(II) and Sn(IV) oxides, whilst Al(III) is evident. In each case, elemental metal peaks are detected, suggesting the nanometer films regardless of various binding energies and escape depths.\cite{Strohmeier1990AnAlloys} The cationic compositions of the passive films were calculated from deconvolution peak fits, are listed in Table \ref{tab:xps-cation-fractions}. Peak fits of the other alloys can be found in the supplementary material.

Cationic enrichment factors ($f$) for each alloy and its constituents Al, Cu, and or Sn were calculated using the equation described by Asami et al.\cite{Asami1978}

\begin{equation}
    f_M = \frac{ {X_M^{film}}/{\Sigma X^{film}_M}}{X_M^{bulk}/ \Sigma X_M^{bulk}}
\end{equation}
where $X^{film}_M$ and $X^{bulk}_M$ are surface (see Table \ref{tab:xps-cation-fractions}) and bulk concentrations (see Table \ref{tab:compositions}) in at.\% of constituent M, respectively. 
Here, $f < 1$ indicates depletion of the element, while $f>1$ indicates enrichment in the passive film relative to the bulk concentrations. For Cu-$x$Al alloys without added Sn, $f_{Al}$ increased from $\approx 0.6$ to $\approx 1.2$ while $f_{Cu}$ decreased from $\approx 1.2$ to $\approx 0.9$ with increasing the Al content in the alloy as shown in Figures \ref{fig:xps}b and \ref{fig:xps}c. Further, alloying Cu-18Al and Cu-23Al with Sn increased $f_{Al}$ from $\approx 0.9$ and $\approx 1.2$ to $\approx 2.25$ for both the alloys, respectively. Further, $f_{Cu}$ also depleted for Cu-18Al and Cu-23Al from $\approx 1$ to $\approx 0.6$ and $\approx 0.9$ to $\approx 0.2$ with Sn addition. Sn enrichment was also observed with $f_{Sn}$ $\approx 30$ for Cu-18Al-0.5Sn, $\approx 20$, and $\approx 12$ for Cu-23Al-1.5Sn and Cu-23Al-2.5Sn alloys, as shown in Figure \ref{fig:xps}d. A significant increase in Al(III) in the passive film is indicated, enhanced by Sn alloying. Figure \ref{fig:xps}e shows the strong correlation between the electrochemical parameter $h$ of the alloy  with $f_{Al}$ obtained via XPS of the passive film after 300 s, all at $+0.6$ V. Values of $h$ decreased with the increasing values of $f_{Al}$. Further, Sn added alloys showed the largest $f_{Al}$ while recording a lower $h$ monolayers required to achieve passivation than Sn free alloys with the same Al content. This indicates that the enrichment of Al(III) and Sn(IV, II) in the passive films during the 300 s of the CA hold allowed the alloy to quickly passivate.

\begin{table}[H]
\centering
\caption{XPS based passive film oxide cation compositions (at.\%) of select Cu-Al-Sn alloys after CA experiments  at $+0.6$ V for 300 s in deaerated \sulfate. }
\label{tab:xps-cation-fractions}
\begin{tabular}{@{}cccc@{}}
\toprule
\textbf{Alloy} &
  \textbf{\begin{tabular}[c]{@{}c@{}}Cu (I, II)\\ (at. \%)\end{tabular}} &
  \textbf{\begin{tabular}[c]{@{}c@{}}Al (III)\\ (at. \%)\end{tabular}} &
  \textbf{\begin{tabular}[c]{@{}c@{}}Sn (II, IV)\\ (at. \%)\end{tabular}} \\ \midrule
\textbf{Cu-16Al}      & 89 & 11 & -  \\
\textbf{Cu-18Al}      & 83 & 17 & -  \\
\textbf{Cu-18Al-0.5Sn} & 46 & 40 & 15 \\
\textbf{Cu-23Al}      & 74 & 26 & -  \\
\textbf{Cu-23Al-1.5Sn} & 19 & 51 & 30 \\
\textbf{Cu-23Al-3Sn}   & 14 & 49 & 37 \\ \bottomrule
\end{tabular}
\end{table}

\subsubsection{Surface Depth Profile}
\begin{figure}[!ht]
    \centering \includegraphics[width=\linewidth]{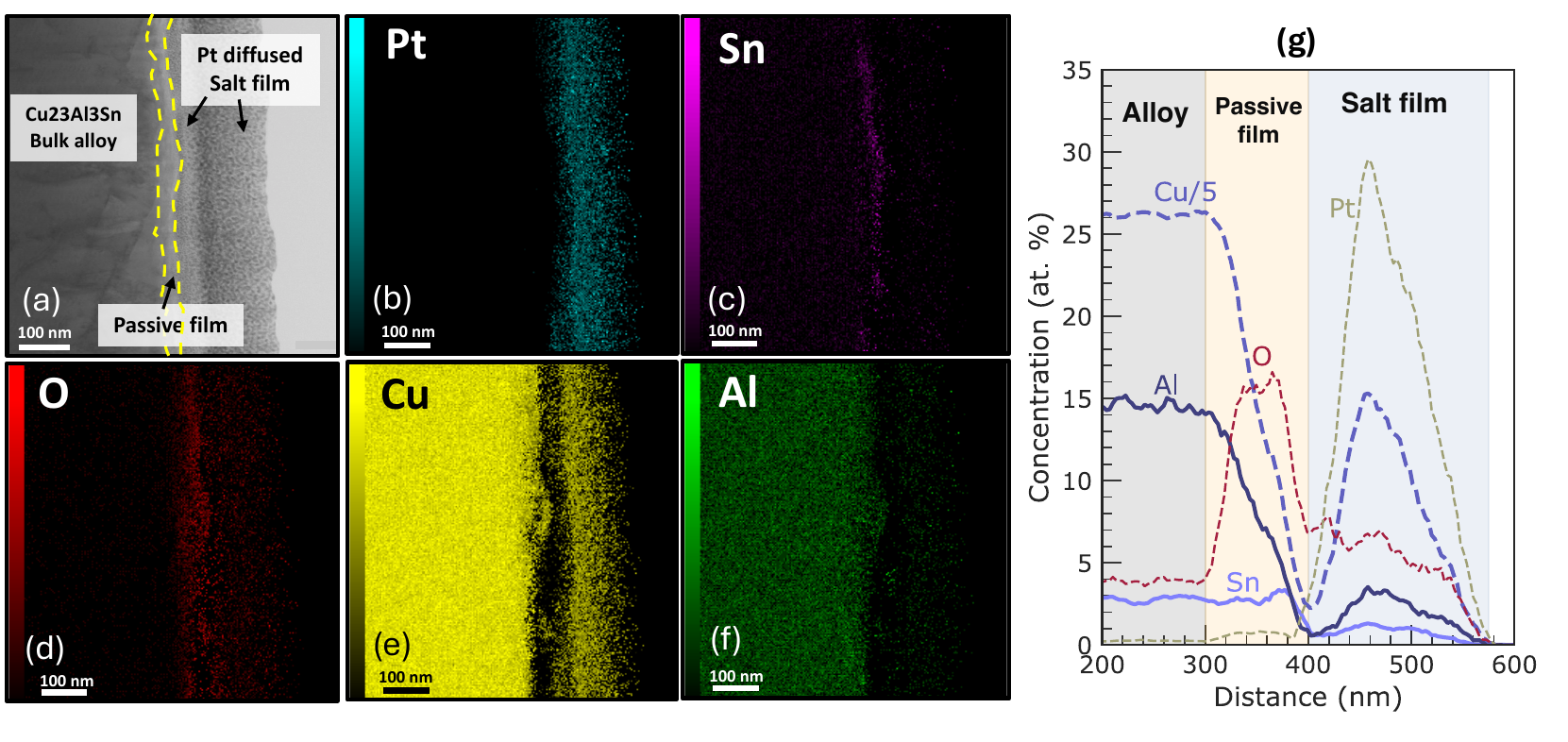}
    \caption{(a) Cross-sectional bright field STEM image, (b-f) STEM-EDS elemental area maps, and (g) concentration profile of Cu, Al, Sn, O, and Pt obtained from a horizontal line scan of the Cu-23Al-3Sn (at.\%) TEM lamellae, post-exposure to single-step CA experiment at +0.6 V for 300 s in deaerated \sulfate. Here, `Cu/5' indicates the concentration of Cu was scaled down by a factor of 5.}
    \label{fig:stem}
\end{figure}

Figure \ref{fig:stem}a-f demonstrates the cross-sectional bright field image and elemental maps near the surface after potentiostatic passive film growth for 300 s at +0.6 V in dearated {\sulfate}. The bright field image shows three distinct zones, including bulk Cu-23Al-3Sn alloy, oxide passive film, and a Pt diffused salt film. Figure \ref{fig:stem}g illustrates a line profile across the region shown in Figure \ref{fig:stem}a. The Pt diffused salt film zone, which is approximately 200 nm thick, is rich in Cu and Al, due to their soluble sulfate salts. The passive film zone, which is rich in O (see the peaks), shows the presence of Sn, Al, and Cu. The STEM/EDS elemental profiles across the passive film corroborate well with their cationic depth profiles obtained via XPS under the same conditions, shown in Figure S11 of the supplementary material. Both indicate that the content of Sn(IV, II) cations is higher in the outer regions of the passive film while oxides of Al (III) and Cu(II, I) are present throughout the passive film zone. This also suggests a cationic mixing of Al(III), Sn(IV), and Cu(II, I) oxides in the passive film.\footnote{Separation of peak heights in the intensity vs. distance or sputter time is assumed to be layered oxides.} However, additional diffraction data is required assuming indication of perhaps a single solid solution oxide or solute trapped oxide.

\section{Discussion}
The main findings herein are that for Cu-$x$Al alloys, (i) Al content controls the passivation in deaerated {\sulfate}, (ii) Sn additions, as a third element in Cu-Al-Sn solid solutions, enrich the Al(III) surface coverage $\Theta_{Al}$ (Figure \ref{fig:AESEC-LSV}) and there is a greater Al(III) enrichment in the passive film (Figure \ref{fig:xps}), 
and (iii) alloying with Sn can result in Al--Al clustering type CSRO in the alloy. This is speculatively correlated with enhanced passive film formation efficiency (Figure \ref{fig:lsv}c, promoting alloy passivation with fewer number of alloy monolayers ($h$) dissolved before passivator islands and nucleated and/or Al-O-Al percolation occurs. Additionally, protection provided from the enrichment of Sn cations themselves in the Al(III) rich passive films of Cu-Al-Sn alloys cannot be ruled out. However, SnO\textsubscript{2} by itself is not protective as an oxide on metallic Sn under the conditions explored (Figure \ref{fig:lsv}a and Figure S5). These results emphasize the TEE of Sn on Al(III), offering another example of designing corrosion-resistant alloys guided by uncovering and taking advantage of potential synergies to boost passivity with a reduced dependency on primary passivators.

\subsection{Thermodynamic factors influencing passivity}
\begin{figure}[H]
    \centering
    \includegraphics[width=0.5\linewidth]{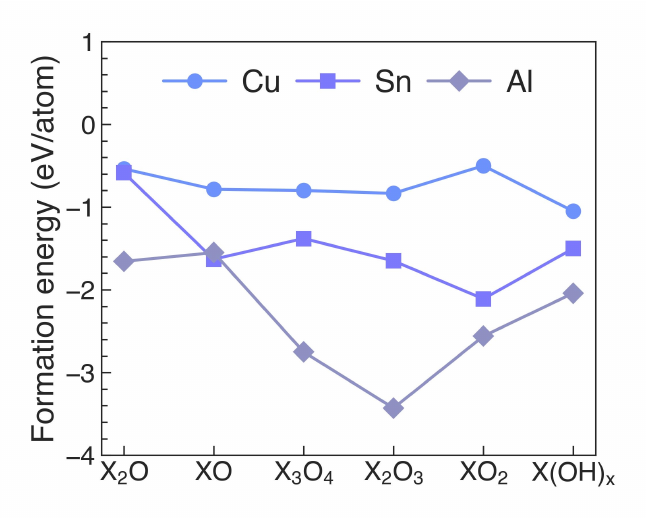}
    \caption{Formation energy of possible oxides and hydroxides of Cu, Al, and Sn extracted from Materials Project at 300 K.\cite{Jain2013}}
    \label{fig:oxides}
\end{figure}

The passive film of an alloy is a mixture of oxides and hydroxides of its constituents, primarily its passivating elements, see Figure \ref{fig:xps}a, which all have high thermodynamic driving forces for oxidation at the applied electrochemical potential of +0.6 V, discussed in this work, see Figure \ref{fig:lsv}a. For Cu-Al-Sn alloys containing more than 18 at.\% Al, Al(III), and Sn(IV, II) cations are the dominant oxidized metallic species observed at $+0.6$ V after 300 s (see Table \ref{tab:xps-cation-fractions}), while Cu undergoes depletion of its Cu(II, I) oxides via chemical dissolution as discussed in Section \ref{sec:aesec}. Thermodynamically, Sn(IV) (tetragonal) and Al(III) (corundum) oxide are the most stable forms with their respective single stoichiometry M-O/[OH]\textsubscript{x} formation energies shown in Figure \ref{fig:oxides}.\cite{Jain2013} Cu(II) hydroxide and oxide are more stable than Cu(I) (cubic) but less stable than Sn(IV, II) and Al(III) oxide. This trend in the most stable oxide correlates well with the surface depth profile configurations of Cu, Al, and Sn cations in the passive film of Cu-23Al-3Sn alloy via STEM-EDS and XPS (see Figures \ref{fig:stem} and S11). However, Al(III) based oxide is more protective than Sn(IV, II) and Cu(II, I) oxides as can be seen from the experimental anodic $i$ values obtained from LSV for high purity Al Sn, and Cu elements in {\sulfate} (see Figure \ref{fig:lsv}a). This indicates that the most stable single metal stoichiometric oxides despite being metastable or unstable at pH 4, are also the most prevalent and protective in this case. The possibility of complex oxides cannot be ruled out. $CuAl_2O_4$ (cubic), $CuAlO_2$ (trigonal), with formation energies $-2.653$ eV/atom and $-2.425$ eV/atom, respectively, have also been experimentally reported. \cite{Jain2013} $AlSnO_3$ (trigonal) with formation energy $-2.279$ eV/atom is also possible but has not been observed experimentally.\cite{Saal2013MaterialsOQMD} However, the complexity of the deconvolution of Cu LMM spectra with several possible oxidation states using available XPS reference spectra and binding energies, along with the difficulty of electron diffraction in thin films, renders fitting and positive identification to many of the complex oxide species mentioned above quite difficult.

\subsection{Kinetic factors affecting passivity}\label{sec: scaffolding}

Beyond oxide thermodynamic stability information, passive film formation is driven by the formation kinetics of the elements and oxides. For instance, $NiO$ with a rock salt structure in Ni-Cr oxides films forms faster than $Cr_2O_3$ (corundum) on FCC Ni-Cr, but over long exposures, becomes more dominant in $Cr_2O_3$ oxide species.\cite{Romanovskaia2024, Lutton2023PassivationSolutions} The very negative $E_{corr}$ of high purity Al suggests that Al(III) oxide should be expected over all applied potentials and even just after cathodic reduction, perhaps before Sn and Cu oxides. High purity Sn can form $SnO_2$ (structure: tetragonal \cite{Jain2013}) with an intermediate step of $Sn(OH)^+$ ion (anodic dissolution) at oxidation potentials below the reversible ion/hydrogen evolution reaction (HER) potential in deaerated {\sulfate}. This transition potential of Sn to $Sn(OH)^+$ ion is also below the $E_{corr}$ of all the Cu-Al-Sn alloys studied in this work (see Figures S5 and \ref{fig:AESEC-LSV}b).

In Sn free Cu-Al alloys, Al dissolved at higher rates than Cu (dealloying) during anodic polarization and showed a significant surface depletion due to its significant difference in between the applied potential and oxidation/reduction potentials (refer to Figure \ref{fig:AESEC-LSV} for Cu-23Al alloy) and chemical instability at pH 4. \cite{Pourbaix1974} In contrast, Cu-23Al-3Sn showed a surface enrichment of Al from AESEC measurements, and XPS peak fitting (see Figures \ref{fig:AESEC-LSV} and \ref{fig:xps}). This stark difference may be due to the apparent benefits of Sn(II) oxide at cathodic and anodic potentials near $E_{corr}$ that can subsequently reduce the net elemental dissolution of Cu(II) and Al(III) over these potentials, leading to surface enrichment of Al and promote the enrichment of more protective Al(III) oxide in the passive film that subsequently forms. It should be noted that Sn (IV, II) oxides are significantly less protective than Al(III) but were observed to be enriched, unlike depleted Cu(II, I) oxy/hydroxides, (see Figures \ref{fig:xps}b-d), suggesting Sn acting as a ``minor passivator" with less dissolution when buried at other roles as discussed below. Herein, a mechanism of ``scaffolding'' provided by Sn(IV, II) oxides of Al(III) oxide during passive film formation is possible, where Sn oxides can allow more protective Al (III) oxides to form while blocking Cu dissolution in the process. This would also align with the TEE provided by Zn in Cu-Al alloys in promoting Al(III) oxide scale formation as $ZnO_2$ acts as an `$O_2$ getter' and reduces oxidation of underlying Cu, allowing Al(III) to form.\cite{Beiwenger1976DasTemperatur, Wagner1965PassivityTemperatures}

\subsection{Influence of chemical short-range ordering}
Another possible mechanism would be the added benefit of clustered Al--Al and potentially Sn--Sn pairs following the percolation theory \cite{Sieradzki1986ASteels, Xie2021AAlloys} based as well as supported enhanced nucleation and growth of oxides based on other passivation models.\cite{Diawara2010NucleationModel, Legrand2002Three-dimensionalAlloys}  The threshold required for Al in single-phase alloys can be as high as 25 at.\% in BCC Fe in 0.0126 M $H_2SO_{4(aq.)}$. \cite{Peng2019} This work suggests that this value can also be considered for Cu-Al alloys and can be verified by testing alloys containing larger Al content. Alloying Sn in Cu does not provide any significant benefit until a large content of more than 10 wt.\% is added.\cite{Mabille2003MechanismConditions, Muller2013-CuSnProcesses} Thus, Al is the superior and primary passivating element in this case. As observed from first principles MC simulations supported by cluster expansion calculations, Al can be induced to cluster via Sn as the clustering agent in their first $nn$ shell, as seen in Figure \ref{fig:csromap}. Here, Al--Al pairs cluster (or exhibit less ordering type for the experimentally studied compositions) which would reduce the Al--Al $nn$ distance, allowing nucleation of Al oxide \cite{Diawara2010NucleationModel, Legrand2002Three-dimensionalAlloys} or percolation of `Al-O-Al' \cite{Sieradzki1986ASteels, Xie2021AAlloys} to occur more rapidly \cite{Sur2025InvestigatingAcid, Blades2024TuningAlloysb, Sur2025FactorsTi}  and achieve faster alloy passivation at the same bulk Al content (see Figures \ref{fig:lsv} and  \ref{fig:AESEC-LSV}).  Further, it should be noted that Sn--Sn pairs significantly also show a tendency of clustering type CSRO with Sn addition, which can promote Sn(IV, II) oxides during the early stages of passivation and foster the \textit{`scaffolding'} mechanism described in Section \ref{sec: scaffolding}. Thus, herein Sn plays a major role as a \textit{`facilitator'} of Al(III) oxide-rich passivity by acting as the agent for introducing clustering type CSRO in Al--Al pairs, decreasing the critical passivity threshold of Al in Cu alloys.

Overall, the synergistic effect of Al and Sn in promoting aqueous passivation can be utilized in designing advanced corrosion-resistant Cu alloys. Moreover, in compositionally complex alloys where chemical short-range order is prominently observed.

\section{Conclusions}
This work investigated the roles of Sn, Al, and Cu in the passivity of arc-melted and solution heat-treated Cu-$x$Al-Sn alloys in deaerated {\sulfate}, using electrochemical parameters such as {\icrit}, {\ipass}, $h$ values, imaginary impedance ($-Z_{imag}$), impedance modulus ($|Z|$) at 5 mHz, spectroelectrochemical parameters including $i_{M}$, and $\Theta_M$, XPS based elemental enrichment factors ($f_M$), and STEM-EDS elemental profiles. The following conclusions were observed that can be used to study further the Al--Sn synergy and design corrosion resistant Cu based alloys:

\begin{itemize}
    \item The addition of small amounts of Sn to Cu-$x$Al alloys where x $>$  18 at.\% substantially enhances the passivation behavior of Cu-$x$Al in a deaerated acidified sulfate solution. 
    
    \item The role of Al was categorized as the primary passivator. Al additions, up to 23 at.\% in Cu demonstrated an increase in corrosion resistance supported by an enrichment of Al(III) cations in the passive film. However, Al additions alone are not enough to achieve excellent corrosion resistance.

    \item With the addition of Sn in Cu-$x$Al ($x>18$  at.\%) alloys, passive film compositions indicated the enrichment of Al(III) and Sn(IV, II) cations along with the presence but depletion of Cu(II, I) from the oxide when expressed as cation fractions. Faster film formation kinetics were found to occur, and an increase in polarization resistance was observed after 10 ks in alloys containing only 18 or 23 at.\% Al with limited Sn additions. These findings indicate a \textit{third element effect} of Sn towards Al(III) in showing stainless type behavior in acidified 0.1 M Na\textsubscript{2}SO\textsubscript{4} solution. Further, Al-Sn synergy was retained with ten times higher polarization resistance when considering Cu-23Al-3Sn compared to Cu-23Al after 28 days of exposure to {\sulfate}.
    
    \item The role of Sn in Cu-Al-Sn alloys was categorized as a `facilitator' as it assists the primary passivator Al in introducing clustering type chemical short-range ordering of Al--Al pairs in an FCC lattice. However, oxidized Sn may lead to forming an Al(III)-Sn(IV, II) mixed oxide that is more protective than Al(III) or Sn(IV, II) after 300 s. Further, Al and Sn enrichment was confirmed during passive film formation using \textit{in-operando} AESEC experiments, suggesting Sn can also be a secondary `minor' passivator. 
\end{itemize}

\section{Acknowledgments}
The authors gratefully acknowledge funding from the Office of Naval Research (ONR) through the Multidisciplinary University Research Initiative (MURI) program (award: N00014-20-1-2368) with program manager Dr. D. Shifler. Thanks to B. Redemann at Johns Hopkins University for casting and solution heat treating two of the alloys, the University of Virginia NMCF for utilizing their PHI Versaprobe III XPS, Bruker D2 XRD, Quanta 650 SEM, and Hirox Light Microscope, and MCP in the Whiting School of Engineering at Johns Hopkins University for utilizing their TEM. D.S. extends his gratitude to the UVA-SEAS Endowed Olsen Graduate Fellowship for partial funding of this work.  J.H. and AESEC experiments conducted by D.S. were also supported by the French government's ``France 2030'' initiative through the PEPR-DIADEM (Priority Research Programs and Equipment - Integrated Devices for Accelerating the Deployment of Emerging Materials) program, managed by the French National Research Agency (Agence Nationale de la Recherche, ANR), n$^\circ$ANR-23-PEXD-0006. E.A.A., D.L. and M.L.T acknowledge partial funding in part from the National Science Foundation MRI program (Award No. 1429661). D.S., J.H., and J.R.S. express their gratitude to Dr. K. Ogle at Chimie Paris Tech for the valuable discussions regarding the AESEC work.

\bibliography{references}

\end{document}